\documentclass[a4paper,abstracton]{scrartcl}

\usepackage[utf8]{inputenc}

\pdfoutput=1

\usepackage{geometry}
\usepackage{amsmath}
\usepackage{amsthm}
\usepackage{amssymb}
\usepackage{appendix}

\usepackage[usenames,dvipsnames]{color}
\usepackage{graphicx}
\usepackage{subfigure}

\usepackage{color}

\usepackage{url}

\usepackage{amssymb,amsmath,verbatim,tabularx,enumitem,colonequals,graphicx,psfrag,subfigure}

\usepackage{cite}

\usepackage{ragged2e}
\usepackage{multirow}

\usepackage{authblk}
\usepackage[normalem]{ulem} 

\usepackage{framed}
\usepackage{rotating}

\newcommand{\BIGOP}[1]{\mathop{\mathchoice%
{\raise-0.22em\hbox{\huge $#1$}}%
{\raise-0.05em\hbox{\Large $#1$}}{\hbox{\large $#1$}}{#1}}}

\allowdisplaybreaks

\usepackage{listings}
\usepackage{xcolor}

\definecolor{codegreen}{rgb}{0,0.6,0}
\definecolor{codegray}{rgb}{0.5,0.5,0.5}
\definecolor{codepurple}{rgb}{0.58,0,0.82}
\definecolor{backcolour}{rgb}{0.95,0.95,0.92}

\lstdefinestyle{mystyle}{
  backgroundcolor=\color{backcolour},   commentstyle=\color{codegreen},
  keywordstyle=\color{magenta},
  numberstyle=\tiny\color{codegray},
  stringstyle=\color{codepurple},
  basicstyle=\ttfamily\footnotesize,
  breakatwhitespace=false,         
  breaklines=true,                 
  captionpos=b,                    
  keepspaces=true,                 
  numbers=left,                    
  numbersep=5pt,                  
  showspaces=false,                
  showstringspaces=false,
  showtabs=false,                  
  tabsize=2
}

\begin{document}

\title{\textcolor{black}{\LARGE Understanding controlled EV charging impacts using scenario-based forecasting models}}



\author[1]{\textbf{\large Rahul Roy}}
\author[2]{\textbf{\large Trivikram Dokka}}
\author[3]{\textbf{\large David A. Ellis}}
\author[4]{\\ \textbf{\large Esther Dudek}}
\author[5]{\textbf{\large Paul Barnfather}}
\affil[1]{\footnotesize Kenan-Flagler Business School, University of North Carolina at Chapel Hill, US, rahulroy@.unc.edu}
\affil[2]{\footnotesize Department of Management Science, Lancaster University, UK, t.dokka@lancaster.ac.uk}
\affil[3]{\footnotesize School of Management, University of Bath, UK, dae30@bath.ac.uk}
\affil[4]{\footnotesize EA Technology, Capenhurst, UK, esther.dudek@eatechnology.com}
\affil[5]{\footnotesize EA Technology, Capenhurst, UK, paul.barnfather@eatechnology.com}

\date{}

\maketitle

\justifying

\abstract{\textcolor{black}{
\textcolor{black}{Electrification of transport is a key strategy in reducing carbon emissions. Many countries have adopted policies of complete but gradual transformation to electric vehicles (EVs). However, mass EV adoption also means a spike in load (kW), which in turn can disrupt existing electricity infrastructure. Smart or controlled charging is widely seen as a potential solution to alleviate this stress on existing networks. While analyses with completely uncontrolled charging result in unrealistically \textcolor{black}{amplified estimates} of EV load, assuming a completely coordinated controlled charging also oversimplifies the estimates of impact due to mass EV uptake. Learning from the recent EV trials in the UK and elsewhere we take into account two key aspects which are largely ignored in current research: EVs actually charging at any given time and wide range of EV types, especially battery capacity-wise. Taking a minimalistic scenario-based approach, we study forecasting models for mean number of active chargers and mean EV consumption for distinct scenarios. Focusing on residential charging the models we consider range from simple regression models to more advanced machine and deep learning models such as XGBoost and LSTMs. We then use these models to evaluate the impacts of different levels of future EV penetration on a specimen distribution transformer that captures typical real-world scenarios. In doing so, we also initiate the study of different types of controlled charging when fully controlled charging is not possible. This aligns with the outcomes from recent trials which show that a sizeable proportion of EV owners may not prefer fully controlled centralized charging. We study two possible control regimes and show that one is more beneficial from load-on-transformer point of view, while the other may be preferred for other objectives. We show that a minimum of 60\% control is required to ensure that transformers are not overloaded during peak hours.} 
}}

\medskip


\section{Introduction}
\label{introduction}
\textcolor{black}{Globally, the EV charging infrastructure has been rapidly expanding to match the growth of EVs}. A study by UK’s National Grid reveals that there would be 90\% penetration of EVs by 2050 leading to an increased \textcolor{black}{energy} demand of 46 TWh; this is over and above the total energy demand of 308 TWh in 2016 \cite{robinson2018electric}. My Electric Avenue (MEA), \textcolor{black}{a 3-year project conducted from 2013 to 2015 in the UK to explore the impact of charging clusters of EVs at peak times on electricity networks}, predicted that approximately 32\% of low-voltage (LV) distribution networks would require intervention when 40-70\% of the vehicles would be EVs \cite{godfrey2016carconnect}. MEA observed that, with rising uptake of EVs, there is a high probability of cluster formation within localities \cite{godfrey2016carconnect}, that is, clusters of consumers with charging requirements occurring at the same time. The additional power demand caused by such clusters would eventually create stress on the distribution networks. Hence, it becomes imperative for the DNOs to have an estimate of the additional power demand from the grids caused solely by EV charging, to ensure seamless demand management, \textcolor{black}{especially during peak hours in a day}, in their local distribution networks. 

\medskip

\textcolor{black}{In this paper, we empirically evaluate the impacts that EV charging would have on local distribution networks using scenario-based forecasting of mean number of EV owners who would charge their EVs per day and the resulting mean energy consumption.}
 
\medskip

\textcolor{black}{The remaining paper is structured as follows. Section \ref{literature} discusses relevant literature on previous researches on EV charging followed by section \ref{contribution} that highlights the research gap and presents our research contribution. Section \ref{data} briefly discusses the EV home-charging data used in this study. Sections \ref{methodology} and \ref{algorithms and results} respectively discuss the analytical methodology, and algorithms and results. We evaluate the impact of EV charging on distribution transformers in section \ref{impact} before concluding this paper in section \ref{conclusions} by summarising our findings and discussing the scope of future research.} 

\section{Related Literature}
\label{literature}
\textcolor{black}{Utilities are interested in both the energy consumption and the additional load on LV networks caused by EV charging. An estimate of the additional energy consumption on the LV distribution networks caused by EV charging is extremely useful for the policy-makers, utilities, and city and district administration councils; \textcolor{black}{for example, in solving unit commitment problem in generation planning and in planning of the roll-out of an EV charging infrastructure in the future.}} Moreover, estimating the additional bulk load requirements prior to the widespread adoption of EV charging is part of emergency preparedness in distribution networks as any additional bulk load on the network may lead to frequent outages or may cause significant damage to the electric infrastructure. \textcolor{black}{It is now widely acknowledged that significant EV uptake will put stress on \textcolor{black}{electricity grids}, prompting many studies to explore impacts of EVs on electricity infrastructure. For example, in their review on the factors relating to plug-in hybrid EVs (PHEVs) that have an impact on distribution networks, \cite{green2011impact} found that driving patterns, charging characteristics, charge timing and vehicle penetration were most relevant. Besides this, \cite{foley2013impacts} studied the impacts of EV charging in an actual working electricity market in Ireland and showed that EV charging had significant impact on wholesale electricity market and off-peak charging was beneficial. While \cite{neaimeh2015probabilistic} used a probabilistic approach, combining EV charging and smart meter household consumption data, to highlight the need for better planning for dealing with stochastic nature of charging demand, \cite{xydas2016data} proposed a ‘risk level’ index using fuzzy logic to assess the impact of EV power demand on distribution networks. Moreover, based on the findings from the Victorian EV trials in Australia, \cite{khoo2014statistical} projected the mean and maximum percentage increase in the \textcolor{black}{power} demands between 3.27\% and 5.70\%, and 5.72\% and 9.79\% respectively in the summer of 2032/33. Based on a test scenario of 57 EVs in an urban locality of Australia, \cite{satarworn2017impact} estimated that a transformer could handle safely up to 75\% EV penetration, provided that the driving performance was not poor. \cite{wamburu2018analyzing} assessed that the share of critically overloaded transformers was low for EV penetration levels in the range of 1-5\% but increased at higher penetration levels, 20-40\% in the New England region of the United States. \textcolor{black}{\cite{gerossier2019modeling} focused on charging stations for 46 privately owned EVs, to analyze and model charging habits for an individual EV owner and estimated that EV charging would have a moderate impact on the electricity network in South Central zone of Texas under an assumed EV market share of 30\% in 2030. \cite{crozier2019impact} also investigated the impact of home charging of private fleet of EVs on power systems. It was estimated that a 100\% penetration of EVs in Great Britain would lead to an increase of 8 GW in demand if the EVs' were assumed to have a battery capacity of 30 kWh.}} 

\medskip

Given its impact on existing infrastructure, naturally, a number of studies focused on predicting either EV energy or power demand using different models and assumptions. For example, \cite{xydas2013electric} implemented data mining methods such as decision tables, decision trees, artificial neural networks and support vector machines, to forecast EV load using data on previous day load, number of the week, day of the week, type of day, number of new plug-ins every half-hour and total charging connections every half-hour. Moreover, \cite{wang2015electric} proposed an offline algorithm based on driving behaviour, road topography information, and traffic situation that gave two energy consumption results, one for the maximum driving speed and the other for the most economical driving speed, to give a first impression to the driver on the possible energy consumption and therefore, the range which the EV can cover even before the actual trip. In addition, \cite{wang2016online} also proposed an online energy consumption algorithm that would help in adjusting the energy consumption prediction during driving of battery EVs (BEVs); this would be based on a number of factors such as vehicle characteristics, driving behaviour, route information, traffic states and weather conditions. While \cite{majidpour2016forecasting} forecast energy consumption for a horizon of 24 hours based on historical data from two different data sets: data from customer charging profile and data from station outlet measurements, \cite{arias2016electric} used historical weather and traffic data to forecast EV charging power demand. Based on the studies of the Korean EV market, \cite{moon2018forecasting} estimated the changes in energy demand based on consumer preferences for EVs, charge time of the day and types of EV supply equipment (EVSE). Total energy demand was estimated using total EV owners, average distance travelled per day, and average fuel efficiency of current EVs. Moving further, \cite{lopez2018demand} proposed a demand response strategy based on machine learning to control EV charging in response to the real-time pricing, such that the overall energy cost of an EV was minimized.

\medskip

It is now widely acknowledged that demand management could play a key role in accommodating the additional bulk load caused by EV charging, thereby leading to researches that focused on developing optimized systems that could minimize the impact of EV charging. For example, \cite{nair2018optimal} proposed an optimized scheduling model which could be helpful in minimizing electricity costs for consumers under EV charging. Besides, \cite{crozier2020opportunity} estimated how smart charging would reduce the extent of network reinforcement from 28\% to 9\% in Great Britain. Furthermore, a demand response mechanism based on the thermal loading of transformers was studied by \cite{pradhan2020reducing} which helped in shifting the EV load to minimize transformer ageing. \textcolor{black}{However, a vital prerequisite prior to implementing any strategy is that DNOs should be equipped to assess how clusters of EVs, when charging under different scenarios, can impact the local distribution networks using readily available information at their disposal in the first place. \textcolor{black}{It is important to note that in several previous studies, \cite{xu2019modal, wu2015electric}, the term \textit{EV consumption} indicated the consumption of energy stored in EV's battery based on a set of features such as EV kinematics, driving data, battery state of charge, and so on. However, in our study, EV consumption refers to the consumption of energy by an EV to charge its battery when connected to the electricity grid.}} Moreover, most of the previous studies primarily focused on charging demand at public charging points. For example, \cite{van2013data} analyzed the actual usage patterns of public charging infrastructure in the city of Amsterdam, based on more than 109,000 charging events in the year 2012-13. \textcolor{black}{However, our focus in this paper is on home EV charging demand, \textcolor{black}{given the consumers' strong preferences of home charging over public charging as reported by \cite{electricnation2019finalreport}.}}

\section{Research Contribution}
\label{contribution}
While it is natural for many factors, as explained and used within the above studies, to affect the EV consumption, typically, most of them are not available to be used within a forecast model and hence, cannot be leveraged by the policy-makers to assess the impact that EV charging at home would have under different future scenarios. For example, information such as driver characteristics including behavioral is hard to obtain. Similarly, route choice and trip details of EV users are unlikely to be available on a continuous basis. Moreover, in most studies, the variability in EV battery capacity was extremely small, thereby failing to capture the actual impact of EV charging under real-world scenarios of EVs with a wide range of battery capacities. As such, we consider a minimalist approach, within a wide array of EV battery capacities, which takes the most basic information such as the EV ownership, day of the week, and season of the year, that is most likely to be available to the DNOs in the future to develop a forecast model that would estimate the consumption of energy from the grid caused by EV charging at home. In addition, we evaluate the impact of EV charging on a specimen distribution transformer at specific levels of controlled charging during peak hours. More specifically, in this paper we are interested in the following questions: 
\begin{enumerate}
    \item \textcolor{black}{What is the expected EV energy consumption at a given ownership level in a given time period? The answer to this question is an important ingredient to the generation level decisions. For example, these forecasts will add into the overall aggregated energy demand and will define the solutions to the unit commitment problem. At the same time they are important for impact assessment at distribution level.}
    \item \textcolor{black}{When a completely controlled charging is possible or assumed then transformers are unlikely to be overloaded due to EV charging, as power drawn by EVs can be fully controlled. On the other hand, when complete control is not possible transformer overloads cannot be ruled out and hence, the question is what level of control is adequate to avoid transformer overloading. In other words,} \textcolor{black}{what is the expected load, especially during peak hours, and how does clusters of EVs, with different characteristics, when charging simultaneously during peak hours, can impact the load on a distribution transformer at different levels of controlled charging, including unconstrained charging? More specifically, we ask this question under two natural control regimes that could be adopted in a managed charging scenario. Such natural control regimes arise from two possible controls: the amount of energy and the number of simultaneous chargers allowed. The second controllable factor, that is, the number of simultaneous chargers depends on the number of EV owners charging on any given day. In the existing literature, to the best of our knowledge, it is assumed that each EV owner charges his/her vehicle everyday. This is contrary to the observed behavior, where in reality, charging frequency strongly relates to the car type and battery size among other things and the number of EVs charging a day varies everyday. In our work we explicitly take this into account by emphasizing the difference between EV owners and \textit{users}, which represents the actual fraction of EV owners who actually charge their EVs per day and generate additional bulk load on the network unlike previous studies where the focus was only on the EV owners.} 
\end{enumerate}  

\textcolor{black}{To answer the above questions,
\begin{enumerate}
    \item  We explore several models which take as an input a future scenario, of total EV ownership and time hereafter referred to as the scenario-based data, to forecast the mean users and the expected EV energy consumption. More specifically, \textcolor{black}{we study statistical models, scalable boosted regression trees (XGBoost), and artificial neural networks (LSTMs)} to forecast mean energy consumption and users, which enables to form an estimate of the (peak) load from the observed data. We emphasize that our focus is forecasting at a more aggregated level of per day compared to some studies which focused on forecasting at higher frequencies of time such as hour.
    \item Using these forecasting models, we study the impact on a specimen distribution transformer, typical of most regions in the UK, within two possible frameworks of centrally controlled charging: (1) consumption control, and (2) user control. In the former case, mean energy consumption of each EV is uniformly controlled during peak hours to ensure identical effect on all the EVs plugged-in for charging, while in the latter case, only a fraction of the EVs is actually controlled during peak hours, allowing the remaining EVs to charge uncontrollably. While the former case ensures impartial control over all the EVs plugged-in for charging, the latter case follows from the findings in \cite{wellington2018report}, where approximately 30\% of EV owners were either undecided or uncomfortable with the idea of centrally controlled charging of EVs. We emphasize that the controlled charging settings we consider are closer to the practical setup and more realistically resemble the real-world scenario unlike most studies which implicitly assume that all EVs are controllable and each EV owner shares full information such as SOC and vehicle availability. 
\end{enumerate}}

\medskip

\textcolor{black}{It is important to note that previous studies on EV charging used features that were region-specific and hence, insights obtained from prior researches could not be generalized. On the contrary, in our study, we used a standard set of features that were not location-dependent, ensuring that our findings would be widely applicable irrespective of the region of application.}

\section{Data}
\label{data}
\textcolor{black}{Data in our study were collected as part of the Electric Nation project \cite{electricnation, electricnation2019finalreport}}, \textcolor{black}{world's largest smart charging trial for EVs at home charging stations}. Data, which contained charging transactions of energy consumption for different EV owners, were collected from February 1, 2017 to December 30, 2018 and had a total of 80,313 observations. The transactions were spread across four stages: uncontrolled and trials 1, 2, and 3. \textcolor{black}{While the uncontrolled stage allowed people to charge their EVs without any constraints on the quantum of power consumed, in trial 1, smart charging was introduced to regulate the EV charging but without informing the consumers. In trial 2, consumers were given mobile applications to enable them to interact with the smart charging system.} Trial 3 observations were biased as the consumers were given incentives to charge their EVs at specific times in a day. The processed data, after removing trial 3 data and non-recoverable, non-imputable missing data caused by technical glitches, consisted of 56,637 observations across 13 variables as shown in table \ref{table:original_vars}, which were then transformed into day-wise time series for analyses and modeling.

\medskip

\textcolor{black}{Note that a different time granularity such as per hour is certainly an interesting and relevant choice. However, such granularity was not suitable for our objective and is more likely a choice in a real-time forecasting scenario and in a data-rich environment.}

\begin{table}[htbp]
    \footnotesize
    \centering
        \begin{tabular}{|l|l|l|c|}
    \hline
    \textbf{Variable} & \textbf{Description of Variable} \\
    \hline
    Charger ID & ID of smarter charger installed at consumer's home\\
    \hline
    Participant ID & ID of consumer participating in trials\\
    \hline
    Car kW & Power rating of battery\\
    \hline
    Car kWh & Energy capacity of battery\\
    \hline
    Group ID & ID of group to which a consumer was assigned during trials\\
    \hline
    Trial & Stage of trials (uncontrolled, 1, 2, or, 3)\\
    \hline
    Adjusted Start Time & Time at which EV was plugged-in to a smart charger\\ 
    \hline
    Adjusted Stop Time & Time at which EV was plugged-out from a smart charger\\ 
    \hline
    Consumed kWh & Energy consumed during EV charging\\ 
    \hline
    Active Charging Start & Time at which EV actually started charging after plug-in\\ 
    \hline
    Car Make & Manufacturer of EV\\ 
    \hline
    Car Model & Model of EV\\ 
    \hline
    EV Type & Type of EV (battery-operated, plug-in hybrid, or range extender)\\ 
    \hline
    \end{tabular}
    \caption{List of variables in EV charging transaction data}
    \label{table:original_vars}
\end{table}

\section{Methodology}
\label{methodology}
\textcolor{black}{Mathematically, we consider the following statistical learning problem for our objective:}

\[y = {f(X)} + \epsilon \label{eq:SLP} \tag{SLP}\] 
\\
$y$ = EV users or consumption\\
$X$ = set of features to forecast $y$\\
$f$ = learning algorithm that maps $y$ and $X$\\
$\epsilon$ = random error, independent of $X$, with mean 0  

\medskip

\textcolor{black}{In this paper, the learning problem is a special case of \ref{eq:SLP}, with $X$ = \{$X_o$, $X_d$, $X_s$\}, where, $X_o$, $X_d$, and, $X_s$ are number of EV owners, day of the week and season of the year respectively.}

\subsection{Clustering of EV Owners}
\label{clustering}
\begin{figure}[htbp]
\begin{center}
   \begin{minipage}{0.5\textwidth}
     \centering
     \includegraphics[width=.9\textwidth]{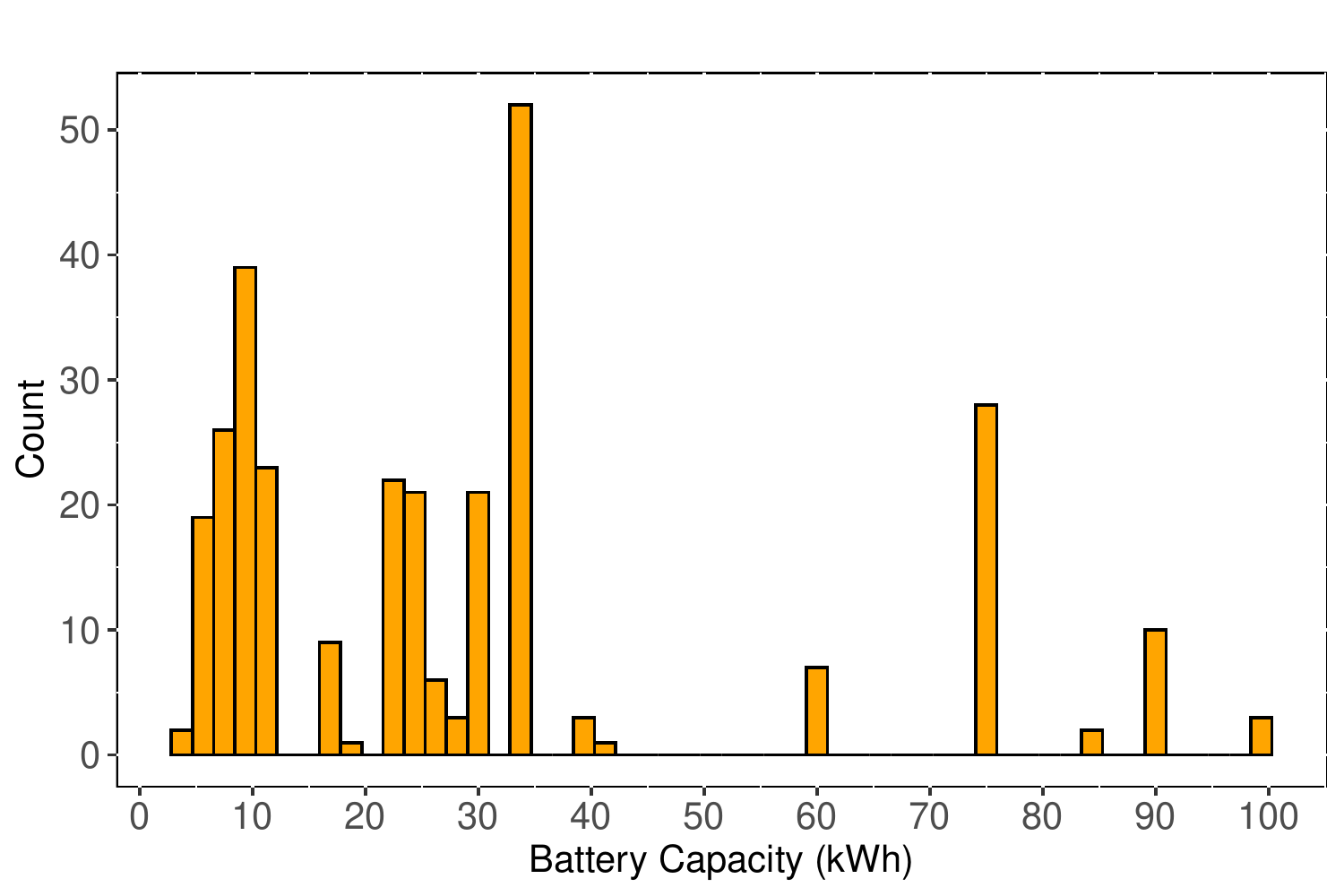}
     \caption{Histogram of battery capacities}
     \label{fig:ratings}
   \end{minipage}\hfill
   \begin{minipage}{0.5\textwidth}
     \centering
     \includegraphics[width=.9\textwidth]{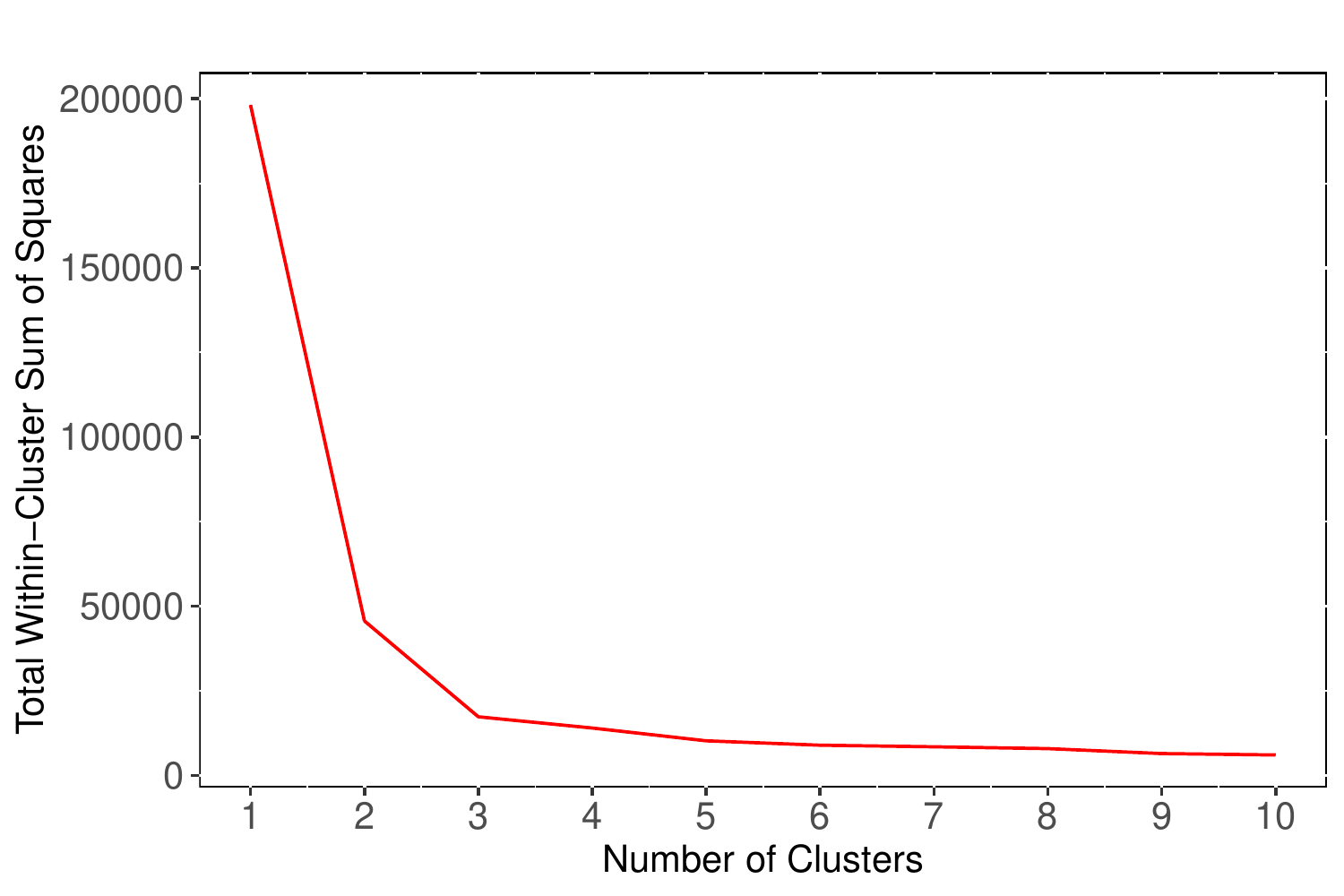}
     \caption{Optimal number of clusters}
     \label{fig:clusters}
   \end{minipage} 
\end{center}
\end{figure}

\textcolor{black}{A total of 26 different battery capacities was present in the data set used. Naturally, each battery capacity combined with car usage pattern could result in a completely different consumption pattern. This implied an important parameter choice in our study: aggregation at battery capacity level. Clearly, building a model for all 26 different battery capacities was impractical. At the same time, ignoring battery capacities and \textcolor{black}{treating all consumption patterns identically} meant ignoring intrinsic variability.} A histogram depicting the variability of battery capacities, \textcolor{black}{caused by the discrete nature of EV battery capacities in the trials}, is shown in figure \ref{fig:ratings}. To validate our assumption of varying charging patterns with distinct battery capacities among EV owners, k-means clustering was performed by transforming the transaction data to obtain a consumer-wise summary of energy consumption per charge against respective battery capacities. Figure~\ref{fig:clusters} shows that based on the consumption behaviour of consumers per transaction and their respective battery capacities, they can be grouped into an optimum number of three clusters as increasing the number of clusters beyond three does not significantly decrease the variability, thereby validating our assumption that one model would fail to capture all the variability of consumer charging behaviour. As mentioned, clustering suggested an optimum number of three clusters. Table~\ref{table:cluster_summary} characterizes the EV owners for all the three clusters. The $Min$ and $Max$ capacities are respectively the minimum and maximum battery capacities of EVs that owners have in the specific clusters. Besides, the \textit{Mean kWh/Charge} is the average energy consumed by an EV owner per charging transaction. \textit{Charging Frequency/Day} illustrates that, on an average, each EV owner does not charge everyday as the charging frequency for each EV owner is always less than one per day. We observe that as the battery capacities increase, the frequency of charging per day decreases and the mean energy consumed per transaction increases, suggesting that with increasing battery capacities, EV owners are less likely to charge per day but when they do, they consume a higher energy per transaction.     

\begin{table}[htbp]
    \footnotesize
    \centering
        \begin{tabular}{|c|c|c|c|c|}
    \hline
   Cluster& Min Capacity & Max Capacity & Mean kWh/Charge & Charging Frequency/Day \\
    \hline
    1 & 4.4 & 18.7 & 5.68 & 0.68\\
    \hline
    2 & 22 & 41 & 14.30 & 0.44\\
    \hline
    3 & 60 & 100 & 26.80 & 0.36\\
    \hline
    \end{tabular}
    \caption{Summary of clusters}
    \label{table:cluster_summary}
\end{table}

\subsection{Time Series Analysis}
\label{time series analysis}
Based on clustering analysis, we transformed the transaction data into three distinct day-wise time series (since we have less than 2 years' data, we do not consider annual but only weekly seasonality in this study; hence, the seasonality of the time series is 7), each belonging to a unique cluster. It is worth mentioning that while transforming the transaction data into day-wise time series, several features were extracted which were not present in the transaction data. Table~\ref{table:variables} lists all the variables (features and target) in the time series. 

\begin{table}[htbp]
    \footnotesize
    \centering
        \begin{tabular}{|l|l|l|c|}
    \hline
    \textbf{Variable} & \textbf{Type} & \textbf{Description of Variable} & \textbf{Notation} \\
    \hline
    day & Feature & Day of transactions & $X_d$\\
    \hline
    season & Feature & Season of the year of transactions & $X_s$\\
    \hline
    owners & Feature & Count of people with EVs per day & $X_o$\\
    \hline
    users & Feature/Target & Count of people charging their EVs per day & $X_u$\\
    \hline
    trans & Feature & Count of transactions per day & $X_{t}$\\
    \hline
    demand & Feature & Upper bound of energy consumed (kWh) per day & $X_{de}$\\
    \hline
    consumed & Target & Total energy consumed (kWh) per day & $y_{c}$\\ 
    \hline
    \end{tabular}
    \caption{List of variables in day-wise time series data}
    \label{table:variables}
\end{table}

All the features except for $owners$ were easily extracted from the transaction data when converting them into day-wise time series. For example, \textit{trans} was the count of all the transactions occurring in a given day, while \textit{users} was the count of all those people who charged their EVs in a given day. However, $owners$ indicated the count of people who owned EVs in a given day. In a real-world scenario, although the count of people with EVs would not change everyday in a distribution network, it would gradually evolve over months and years. However, in Electric Nation project, \textcolor{black}{the number of participants joining the EV trials increased steadily at a much higher rate than that in case of real-world scenario}, resulting in the count of people with EVs changing almost everyday. In a real-world set up, DNOs would not have the information on the actual number of people who would be charging their vehicles everyday as it's a random variable that would depend on a lot of factors and hence, needs to be forecast; for example, day of the week, battery state of charge (SOC), and so on, and would almost always be less than the number of EV owners in a network (figure \ref{fig:owners}). However, \textcolor{black}{based on the charge-point installation notifications, the DNOs would have an estimate of the} number of people in their network who own EVs and hence, knowledge of EV owners is essential to the objective. Besides owners, DNOs would have information on season of the year and day of the week too. In a nutshell, the information available to the DNOs would most likely contain EV owners, day of the week and season of the year. Any additional information available to the DNOs is uncertain and hence, our objective was to develop a forecast model that could be leveraged by the DNOs to forecast both EV users and their energy consumption based on the minimal available information. Since the number of owners could not be directly extracted from the transaction data, we made a few assumptions to compute the number of owners: (1) an EV owner joined the trials whenever he (she) charged his (her) EV for the first time; (2) an EV owner, after signing up, never dropped out of the trials until the trials were inactive. Under these assumptions, the count of EV owners increased with time during the trials.

\medskip

In table \ref{table:variables}, $demand$, or the upper bound of energy consumed (kWh) per day is given by \ref{eq:LTR}.

\[X_{de} = \sum\limits_{i} b_{i} n_{i} \hspace{0.5em} \forall i \label{eq:LTR} \tag{LTR}\]
\\
$b_{i}$ = capacity of $ith$ battery\\
$n_i$ = number of transactions per day for $b_{i}$

\medskip

The three time series generated from the transaction data had missing days between different pairs of dates. To impute the data for those missing days, we adopted a three-step method as discussed in \cite{hyndman2018forecasting}: (1) \textcolor{black}{STL decomposition \cite{cleveland1990stl} was computed to obtain seasonally adjusted data}; (2) linear interpolation was then carried out for the seasonally adjusted data; (3) seasonal component was added back to the linearly interpolated data. Besides imputing the missing values, few observations across the three time series were identified as outliers and replaced with suitable values via a two-fold approach: (1) periodic STL decomposition was carried out to identify observations that seemed unusual from rest of the observations \cite{hyndman2018forecasting}; (2) time plots were analyzed to identify sudden changes in the values that, in turn, helped in identifying which observations were apparently unusual from the rest. The three time series with clusters 1, 2 and 3 had respectively 572, 573 and 573 observations after the final phase of data processing.

\subsection{The Nested Modeling Approach}
\label{nested modeling}
Depending upon what is assumed to be known when forecasting, we can classify forecasts into three categories \cite{hyndman2018forecasting}: (1) ex-ante (to forecast target, we need to forecast features as no information is available on the future values of features); (2) ex-post (information on features is available prior to forecasting); (3) scenario-based (possible scenarios for the features that are of interest to the objective are considered). \textcolor{black}{In this study, scenario-based forecasting is the obvious choice}. 

\begin{figure}[htbp]
\begin{center}
\subfigure[EV owners \label{fig:owners}]{\includegraphics[width=0.45\linewidth]{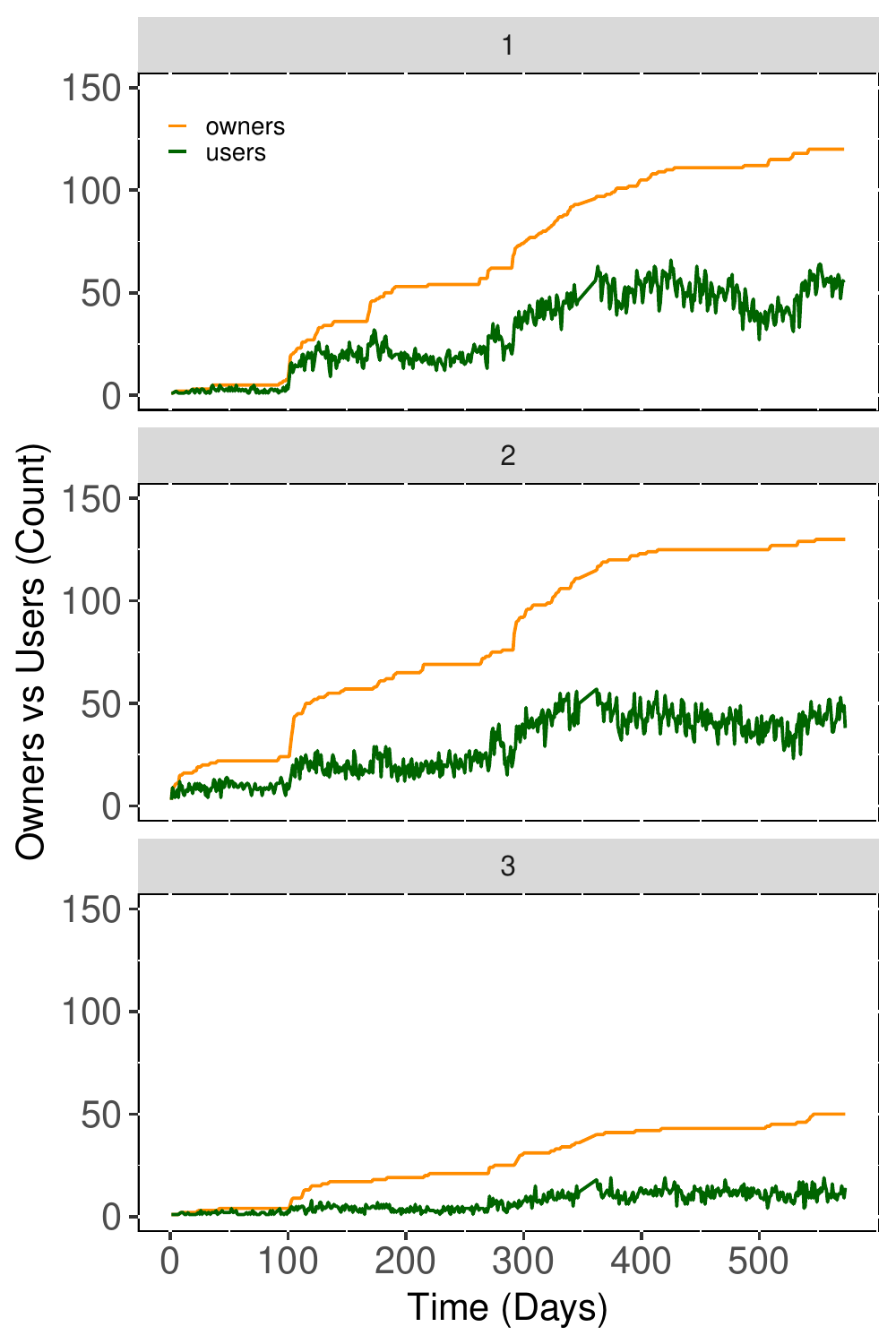}}
\subfigure[EV consumption (kWh) \label{fig:kwh}]{\includegraphics[width=0.45\linewidth]{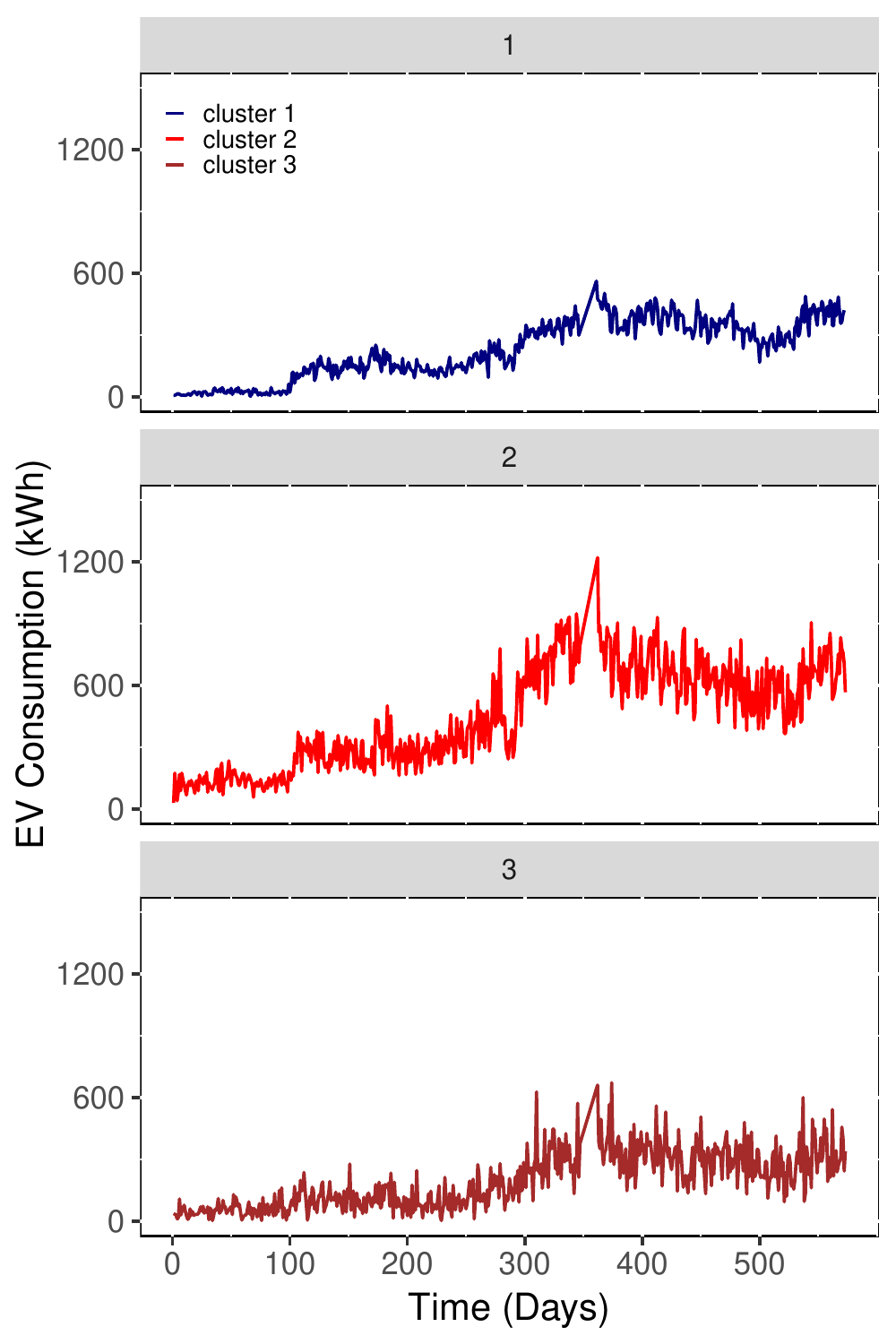}}
\caption{Variation of EV owners, users, and consumption}
\label{fig:timeplot}
\end{center}
\end{figure}

Furthermore, classical univariate methods such as exponential smoothing and ARIMA were not applicable in our scenario-based forecasting objective for the following reasons.

\begin{itemize}
    \item The trend (direction in which a time series slopes) in users and consumption of energy were governed by the number of EV owners in the trials, which itself was a controlled variable. Hence, the trend of users and energy consumption during the trials did not represent the real-world scenario. Figure \ref{fig:timeplot} shows the upward trend of owners and users (figure \ref{fig:owners}) and consumption of energy (figure \ref{fig:kwh}), indicating that as more people joined the trials everyday, the number of users and consumption of energy increased. A univariate method would forecast based on the trend captured during the trials and hence, would result in inaccurate forecasts at a specified period of time in the future. 
    \item If a univariate method was used for forecasting, we would deviate from the objective of scenario-based forecasting as the forecasts would correspond to a specific combination of $X_{o}$, $X_{d}$, and $X_{s}$ at a fixed time stamp in future. In a nutshell, the univariate methods would not be able to generate forecasts for any combination of $X_{o}$, $X_{d}$, and $X_{s}$ based on a decision maker's choice.  
\end{itemize}


\textcolor{black}{\ref{eq:SLP} shows that the objective, involving scenario-based data, does not avail information from all the features in the time series data, that is, while the time series data have six features, the data in the objective involves only three.} 

\begin{figure}[htbp]
\begin{center}
\subfigure[Correlogram (4.4 - 18.7 kWh) \label{fig:corr_c1}]{\includegraphics[width=0.45\linewidth]{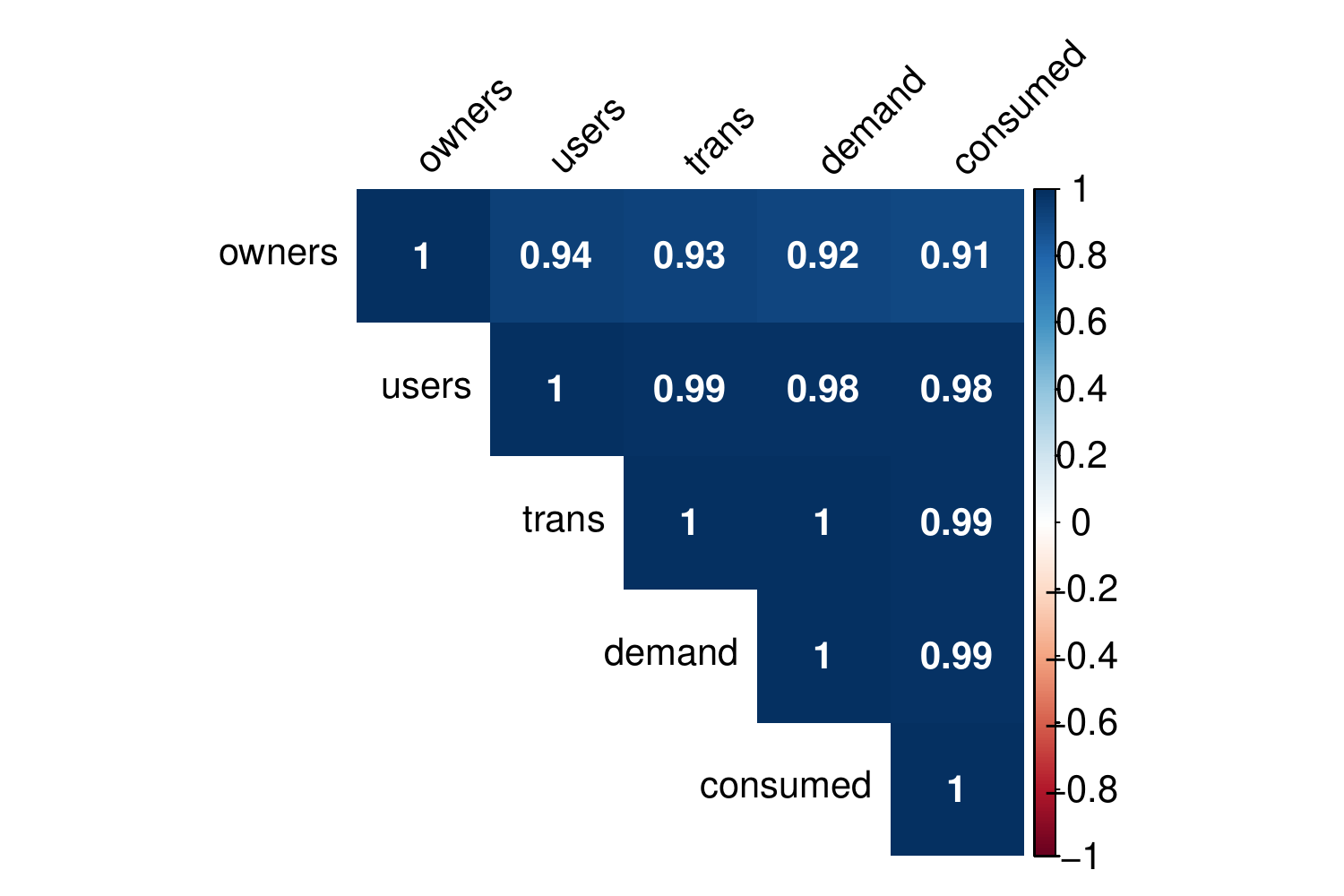}}
\subfigure[Correlogram (22 - 41 kWh) \label{fig:corr_c2}]{\includegraphics[width=0.45\linewidth]{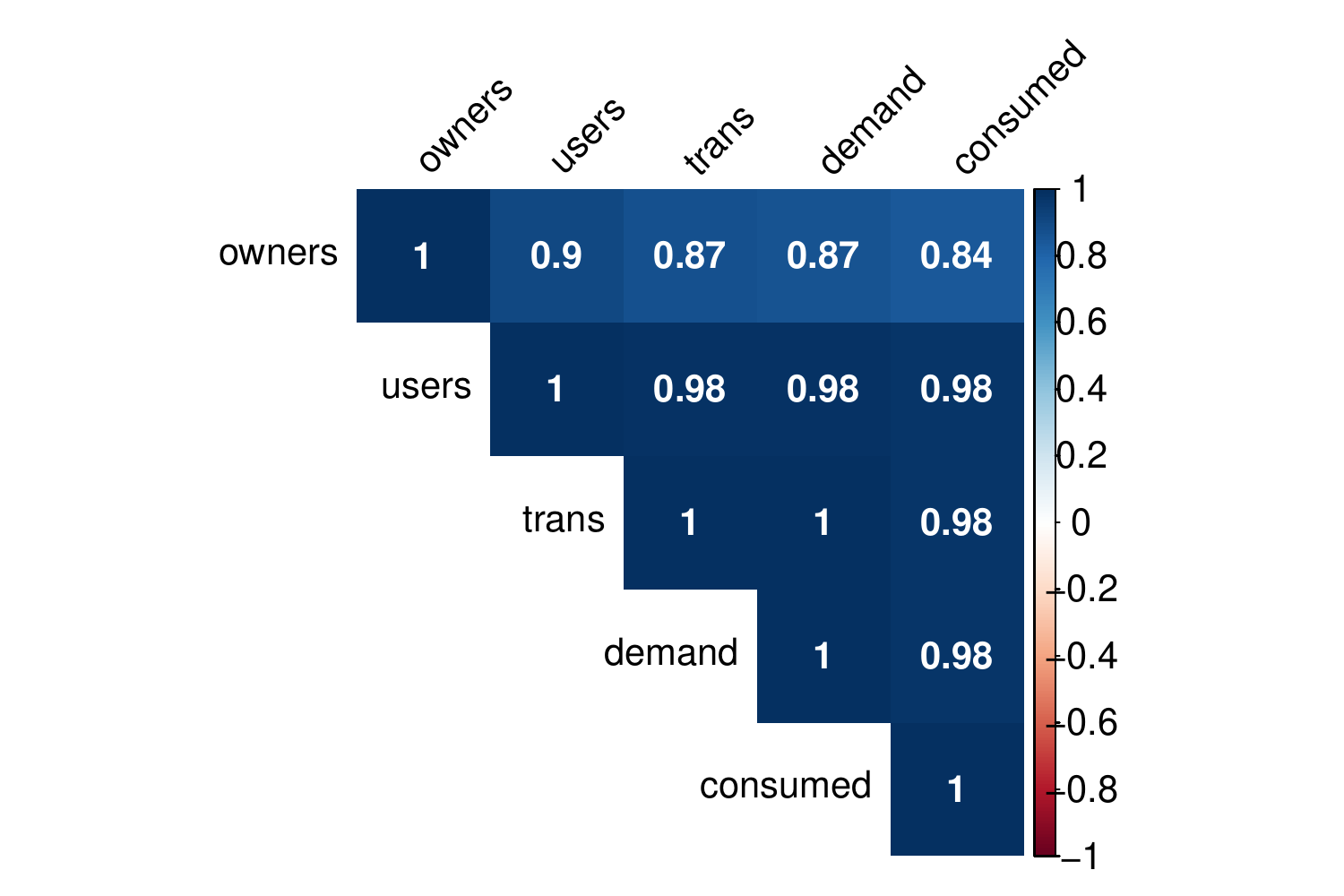}}
\subfigure[Correlogram (60 - 100 kWh) \label{fig:corr_c3}]{\includegraphics[width=0.45\linewidth]{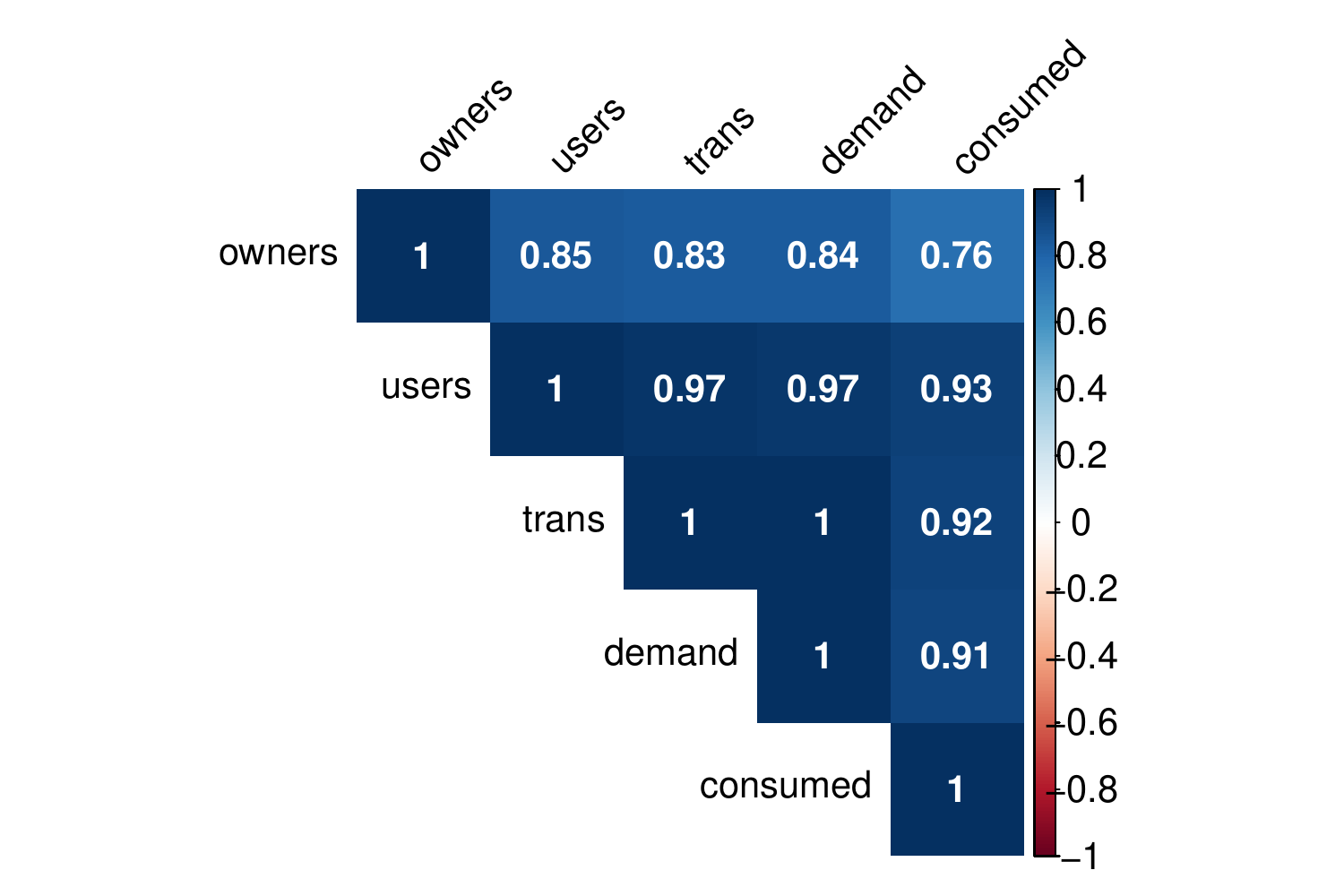}}
\caption{Correlogram of numeric variables}
\label{fig:correlogram}
\end{center}
\end{figure}

\textcolor{black}{Besides this, Figure \ref{fig:correlogram} enumerates the high values of the Pearson's correlation coefficients among all the numeric variables. This indicates that features other than $X_{o}$ might also have an effect on EV users and consumption, thereby necessitating further inspection.} \textcolor{black}{In addition, this also corroborates our assumption that if we use only those features from the time series data that are present in the scenario-based data to fit a model, we fail to make use of the additional information present in the time series data, adversely affecting the forecasting performance. Hence, to ensure that we extract maximal information from the time series data while addressing the mismatch between the feature spaces of the aforementioned data sets, we \textcolor{black}{evaluate} a \textit{nested modeling} approach to forecast EV consumption (Appendix \ref{nested_plot}: figure \ref{fig:nested}) as explained below. \textcolor{black}{It is worth noting that the nested modeling approach cannot be leveraged to forecast EV users as all variables except for the ones in the scenario-based data are a consequence of users but not the other way round. Hence, the only set of features that is available to forecast users comprises of EV owners, day, and season.}}

\begin{itemize}
    \item \textcolor{black}{In the real-world set-up, forecasts will be generated using the scenario-based data; this means that in our modeling framework, our test sample should be similar to the scenario-based data. To begin with, we firstly split the time series data into training and test samples and drop all the features from the test sample except for $X_{o}$, $X_{d}$, and $X_{s}$, to generate a truncated test sample; this ensures that while the training sample resembles the time series data in feature space, the truncated test sample, with its three features, is equivalent to the scenario-based data of the DNOs. However, the target variable, EV consumption, is retained in both the samples.} 
    \item \textcolor{black}{Since DNOs will need to generate forecasts using scenario-based data (equivalent to the truncated test sample) but we also need to leverage the additional information available in the training sample (equivalent to time series data), we affix new features, called \textit{pseudo features} or $p-features$ (p stands for pseudo), to the truncated test sample, to obtain a modified test sample; these p-features are actually the forecasts of the those original features, called $o-features$, which are dropped from the test sample after the train-test split. A p-feature is obtained by: (1) firstly, fitting a model of an o-feature ($X_{j}$), corresponding to a p-feature ($p-X_{j}$), using $X_{o}$, $X_{d}$, and $X_{s}$ from the training sample; (2) secondly, generating forecasts of $X_{j}$ using $X_{o}$, $X_{d}$, and $X_{s}$ from the truncated test sample, and appending them as $p-X_{j}$ t$X_{j}$o the truncated test sample to obtain the modified test sample. For example, we can add a p-feature of $X_{u}$, called $p-X_{u}$, by firstly fitting a model of $X_{u}$ using $X_{o}$, $X_{d}$, and $X_{s}$ from the training sample and then, forecasting $X_{u}$ using the same features from the truncated test sample, to obtain $p-X_{u}$. \textcolor{black}{We repeat the step for all $X_{j}$ except for $X_{o}$, $X_{d}$, and $X_{s}$ in the training sample, to generate a modified test sample that contains $X_{o}$, $X_{d}$, and $X_{s}$ and $p-X_{j}$ corresponding to all $X_{j}$.}}
    \item \textcolor{black}{After we obtain the modified test sample containing all $p-X_{j}$, we again fit a model of an o-feature (target o-feature) ($X_{q}$) using a different o-feature (input o-feature) ($X_{j}$) and generate another set of forecasts of the target o-feature, $p-X_{q'}$. We compare the MAPEs of $p-X_{q'}$ and $p-X_{q}$ (already present in the modified test sample), to identify which forecasts yield low error. If MAPE of $p-X_{q'}$ is found to be less than MAPE of $p-X_{q}$, we assign $p-X_{q'}$ to $p-X_{q}$ to replace $p-X_{q}$ with new values. If MAPE of $p-X_{q'}$ is found to be greater than MAPE of $p-X_{q}$, we reject $p-X_{q'}$. We repeat the step for all $X_{q}$ to obtain the final modified test sample, keeping in mind the causality among features. Figures \ref{fig:nm_1} and \ref{fig:nm_2} illustrate all the aforementioned steps.}
    \item \textcolor{black}{We now use the training sample to fit a model for EV consumption using $X_{j}$ corresponding to $p-X_{j}$. We then use $p-X_{j}$ from the final modified test sample to compute the forecasts for EV consumption as illustrated in figure \ref{fig:nm_3}. In the real-world set-up, DNOs will utilize the scenario-based data as test sample to forecast and then append p-features to it using the models of o-features trained on the complete time series data of the trials as explained in previous steps. Once the final modified version of the scenario-based data is obtained, EV consumption forecasts would be generated using the EV consumption models trained on the complete time series data as explained earlier.}
\end{itemize}

\textcolor{black}{We observe from figures \ref{fig:corr_c1}, \ref{fig:corr_c2}, and \ref{fig:corr_c3} that correlation coefficient between $X_{t}$ and $X_{de}$ is 1, that is, they are perfectly correlated for all the three clusters. This is because $X_{de}$ is obtained by linearly transforming $X_{t}$ as shown in \ref{eq:LTR}. However, $X_{de}$ also includes information on the EV battery capacities and hence, might be influential in forecasting EV consumption as battery capacity is an indicator of EV's consumption capacity.}



\medskip

\textcolor{black}{We call this approach \textit{nested modeling} as we repeatedly fit auxiliary models and generate forecasts internally within a nest-like loop before eventually forecasting EV consumption.}

\subsection{Evaluation on Variable Origin}
\label{variable origin}
\textcolor{black}{The choice of train-test split is user-specific, and it can not be ascertained that a given split is better than the other. To obviate any bias due to a specific train:test split, we evaluate the forecasting performance on a $variable$ $origin$}. In performance evaluation on a variable origin, we firstly fit a model on the first 70\% of the data (training sample) and then evaluate the performance on the last 30\% of the data (test sample). Subsequently, we increase the training sample to 80\% and 90\% of the data and fit models on these samples. Model performances are then evaluated on test samples comprising of the last 20\% and 10\% of the data respectively. The final performance is the mean of all the three performances. We choose the mean absolute percentage error (\ref{eq:mape}) as the error metric for performance evaluation.

\[MAPE = \frac{1}{n} \sum\limits_{i = 1}^{n} \frac{|y_i - \hat{y_i}|}{|y_i|} \label{eq:mape} \tag{MAPE}\] 


\section{Algorithms and Results}
\label{algorithms and results}
In this section, we discuss four algorithms to solve \ref{eq:SLP}. \textcolor{black}{Before fitting a model, data were normalized using the \textit{min-max} scaling.} Moreover, we also present a quantitative comparison of the performances of the final forecast models.

\subsection{Time Series (TS) Regression}
\label{regression}
In TS regression, a dependent variable is modeled as a weighted linear sum of the independent variables, where all the variables are time series. \textcolor{black}{In our case, we set time series regression as the benchmark algorithm, where we modeled EV users or consumption, $y_t$, as a weighted linear combination of features, $x_t$, where the weights are the regression coefficients}. For example, the equation \(y_{ct} = \beta_0 + \beta_1x_{ot} + \beta_2x_{dt} + \beta_3x_{st} + \epsilon_t\) models $y_{c}$ as a function of $X_{o}$, $X_{d}$, and $X_{s}$, with $\epsilon_t$ being the regression error. 

\subsection{Regression with ARIMA Errors (reg-ARIMA)}
\label{reg-arima}
In reg-ARIMA, an ARIMA model is fit on the errors of a regression model and the forecasts from both the regression and the ARIMA components are combined. It is particularly useful when the regression errors show high correlations among each other, indicating that the regression model does not capture all the information in the data. Mathematically, a reg-ARIMA model to forecast $y_t$ as a function $x_t$ can be given by: 
\[y_{t} = \beta_0 + \beta_1x_t + \eta_t\] \[\eta_t = \phi\eta_{t-1} + \epsilon_t + \theta\epsilon_{t-1}\]
As mentioned earlier, an ARIMA model is fitted on the regression errors $\eta_t$. Here, $\phi$ and $\theta$ represent the model parameters corresponding to the AR and MA components of the ARIMA model, while $\epsilon_t$ is the ARIMA error. \textcolor{black}{In this paper, we implemented the Hyndman-Khandakar algorithm \cite{hyndman2007automatic} to tune the orders of the AR and MA components and subsequently fitted the ARIMA model on the regression error, which was obtained after fitting a regression model as explained in section \ref{regression}}.

\subsection{Extreme Gradient Boosting (XGB)}
\label{xgb}
\textcolor{black}{XGB is a scalable, tree boosting system that leverages the \textit{Gradient Boosting} \cite{friedman2001greedy} framework to learn from data and offers superior computational efficiency over other acclaimed machine learning algorithms \cite{xgboost}. The popularity of XGB can be ascertained from the fact that in 2015, 17 out of 29 challenge winning solutions at the machine learning competition site $Kaggle$ used XGB \cite{chen2016xgboost}.}

\medskip

\begin{figure}[htbp]
\begin{center}
\subfigure[Time-slicing (first iteration) \label{fig:timeslice1}]{\includegraphics[width=0.45\linewidth]{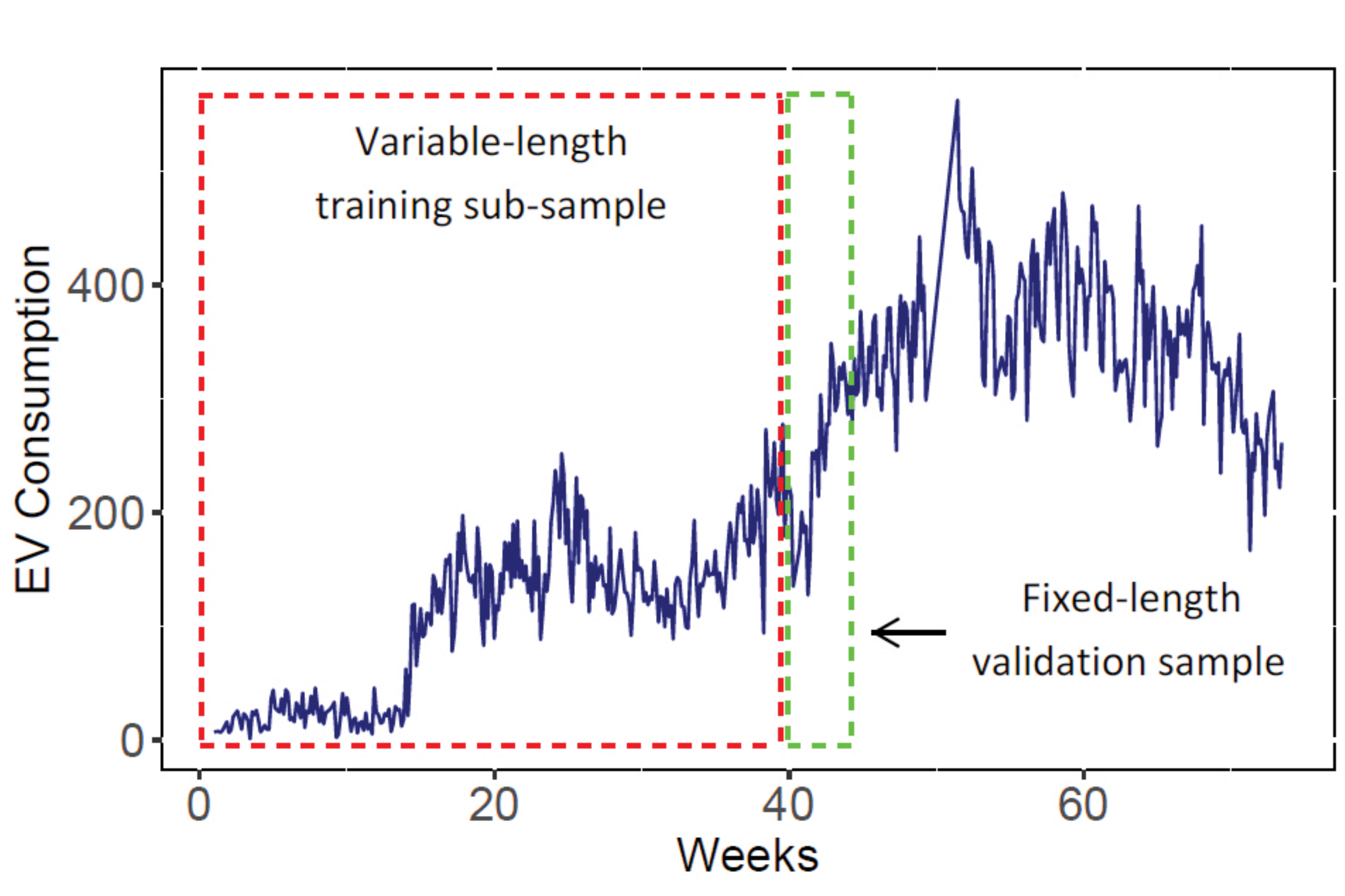}}
\subfigure[Time-slicing (N iterations) \label{fig:timeslice2}]{\includegraphics[width=0.45\linewidth]{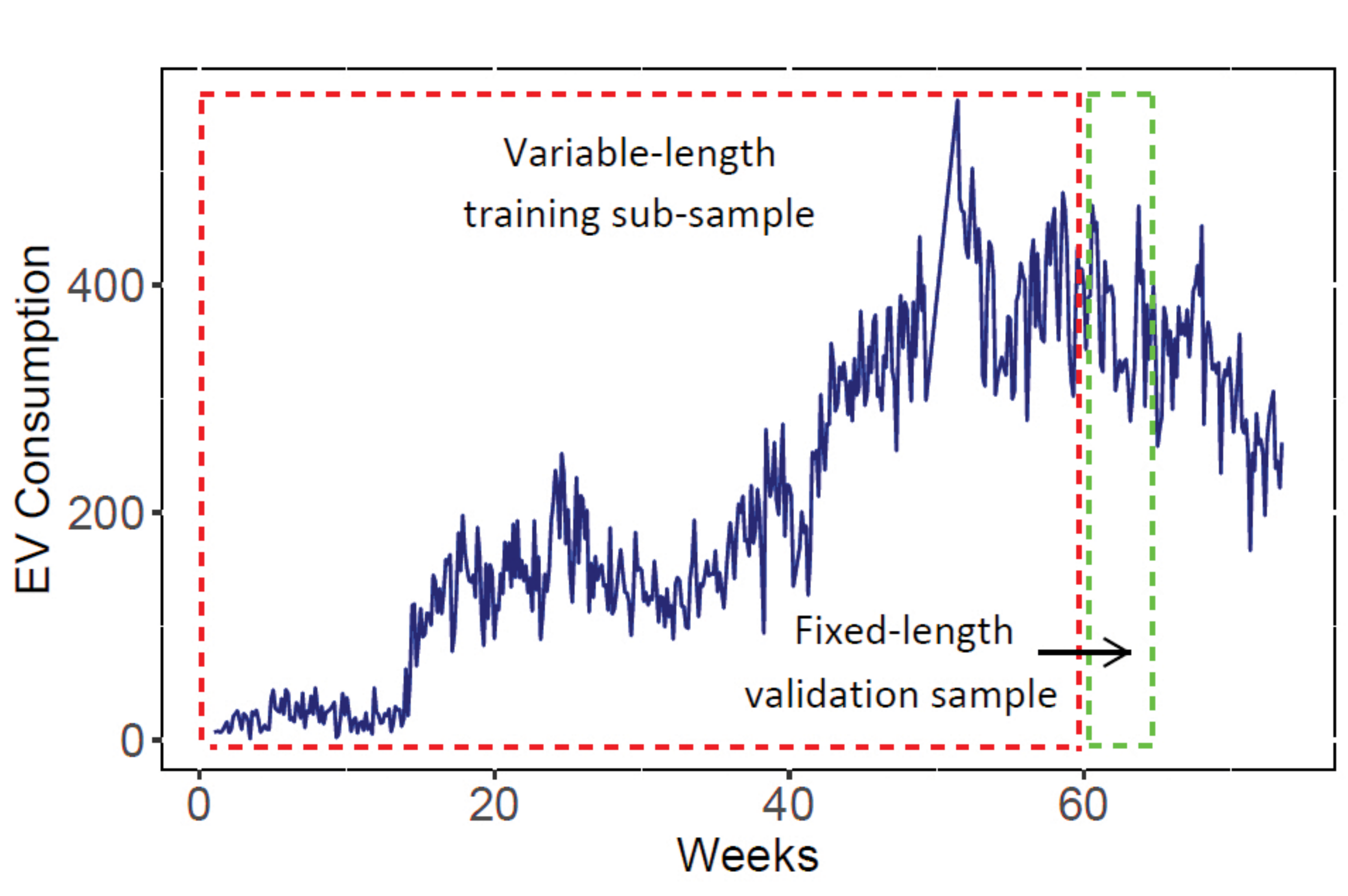}}
\caption{Time-slicing for hyperparameter tuning of XGB}
\label{fig:timeslicing}
\end{center}
\end{figure}

\textcolor{black}{Given the computational resources, we tested between 108 combinations of hyperparameters for each iteration of evaluation on a variable origin via random search in the hyperparameter space. Since our data was sequential, we could not use cross-validation via random sub-sampling to tune the hyperparameters as it would have disarrayed the time dynamics of the data. Instead, we implemented time-slicing to create sub-samples of the training data into a variable-length training sub-sample and a fixed-length validation sample, and iteratively increased the size of the training sub-sample by 1 seasonal difference. Figures \ref{fig:timeslice1} and \ref{fig:timeslice2} (figure \ref{fig:timeslicing}) show time-slicing for hyperparameter tuning in XGB.} 

\subsection{LSTM Networks}
\label{lstm}
\textcolor{black}{LSTMs \cite{hochreiter1997long} are an enhanced version of \textit{Recurrent Neural Networks} (RNNs) which overcome a major limitation faced by the conventional RNNs: \textit{Vanishing Gradient Problem}, in which a network fails to learn long-term dependencies \cite{hochreiter1998vanishing}. In a sequence prediction problem, learning long-term temporal dependencies along with the present state of the system is essential in predicting the future state. LSTMs, courtesy the gating mechanism inside their specially architectured memory cells, regulate the flux of information and enforce a constant error flow through the network, thereby obviating the complications of \textit{vanishing} and \textit{exploding} gradients and enabling the LSTMs to capture long-term temporal dependencies. LSTMs' ability to learn such long-term dependencies have piqued interests among researchers to leverage it for a plethora of sequence prediction problems. For example, \cite{bandara2019sales} used a special variant of LSTMs, known as LSTMs with peephole connections, to forecast sales demand in e-commerce. Moreover, \cite{du2018time} proposed a sequence-to-sequence deep learning framework based on LSTM encoder-decoder architecture for multivariate time series forecasting on air quality data. Besides this, \cite{9107249} used the LSTM-based encoder-decoder architecture to predict long-term traffic flows. Furthermore, \cite{sehovac2020deep} assessed attention mechanisms with different types of RNN cells (vanilla, LSTM, and GRU) and forecasting horizons, for electrical load forecasting.}

\medskip

\textcolor{black}{In this study, we compared a suit of network architectures: $vanilla$ (one hidden layer between input and output layers) vs $stacked$ (multiple hidden layers between input and output layers), $unidirectional$ vs $bidirectional$ \cite{schuster1997bidirectional}, and $vector-output$ vs $encoder-decoder$ \cite{cho2014learning, sutskever2014sequence}, to assess the effect on forecasting performance. It is important to note that although both vector-output and encoder-decoder architectures are popular choices when forecasting horizon ($h$) is greater than 1, we evaluated them in our study as a special case for $h$ being 1. Leveraging these 2 architectures ensures that, with nominal transformations to the input data and output layers of the networks, the LSTMs are flexible to output forecasts of any horizon, 1 or higher, depending upon user requirements. While designing the networks, we set the maximum number of hidden layers to 2, that is, the depth of LSTMs would be either 1 (vanilla) or 2 (stacked) layers. This implies that in case of vector-output architecture, the maximum number of LSTM layers cannot exceed 2. In addition, as encoder-decoder architectures have 2 components, encoder and decoder, where each component is an LSTM network, we specifically assessed the effect of the encoder depth (vanilla or stacked) on the forecasting performance by tuning the decoder depth for a given encoder depth. Besides, the number of neurons in each hidden layer was tuned between 50 and 200. We chose a batch size of 7, \textit{mean squared error} as the cost function, and \textit{Adam} \cite{kingma2014adam} as the optimization algorithm to minimize the cost function, with learning rate ranging from 0.0001 to 0.01 on a logarithmic scale. In addition, we also introduced regularization via neuron dropouts \cite{hinton2012improving}, ranging from 0 (no dropout) to 0.4 (randomly removing 40\% of neurons in each iteration of training), to avoid over-fitting. To tune the hyperparameteres, we used \textit{Bayesian Optimization} \cite{snoek2012practical}. It is important to note that given the computational resources, we set the maximum number of search iterations in Bayesian optimization to 10 and the maximum number of training epochs to 100 for tuning the hyperparameters.}  

\subsection{Results}
\label{results}
\textcolor{black}{Table \ref{table:mape_users} enumerates the mean MAPE values on a variable origin of forecasts of EV users using the features in the scenario-based data,} \textcolor{black}{while table \ref{table:mape_ev} lists the mean MAPE values on a variable origin of EV consumption forecasts for each feature in the data set across all the four algorithms and three clusters. The MAPE values of p-features are tabulated in Appendix \ref{p-features}. We observe that LSTMs deliver the lowest MAPE values among all the algorithms concerned, although XGB performed similar to LSTMs for clusters 2 and 3 for EV users and consumption forecasts and hence, can be preferred to LSTMs if computational budget is a significant consideration.} 

\medskip

\textcolor{black}{We note that out of the twelve best models (four algorithms $\times$ three clusters), six (three for time series regression, two for reg-ARIMA, and one for XGB) use p-features to deliver the best forecasting performance. On investigating the performance of the two best algorithms, XGB and LSTMs, we further observe that to deliver the lowest MAPE values, while XGB uses the p-feature, $p-X_{de}$, in cluster 1 but $X_{o}$, $X_{d}$, and $X_{s}$ in clusters 2 and 3, LSTMs never use the p-features. However, it is worth mentioning that the XGB and LSTM results are constrained by the number of search iterations during hyperparameter tuning and hence, the possibility of obtaining a better forecasting performance with p-features for all or some of the clusters can not be ruled out, thereby necessitating that relevant p-features should be appended to the scenario-based data of the DNOs before forecasting EV consumption, using nested modeling as discussed in section \ref{nested modeling}.}

\begin{table}[htbp]
    \footnotesize
    \centering
        \begin{tabular}{|c|c|c|c|c|c|}
    \hline
    \textbf{Cluster} & \textbf{Feature (s) for Forecasting} & \textbf{Regression} & \textbf{reg-ARIMA} & \textbf{XGB} & \textbf{LSTMs}\\
    \hline
    \multirow{1}{*}{1} & $X_{o}$, $X_{d}$, $X_{s}$ & 19.38 & 22 & 18.55 & 12.01\\
    \hline
    \multirow{1}{*}{2} & $X_{o}$, $X_{d}$, $X_{s}$ & 21.99 & 14.42 & 13.6 & 11.44\\
    \hline
    \multirow{1}{*}{3} & $X_{o}$, $X_{d}$, $X_{s}$ & 29.42 & 28.55 & 23.93 & 23.87\\
    \hline
    \end{tabular}
    \caption{MAPE of EV users forecasts for all clusters}
    \label{table:mape_users}
\end{table}

\begin{table}[htbp]
    \footnotesize
    \centering
        \begin{tabular}{|c|c|c|c|c|c|}
    \hline
    \textbf{Cluster} & \textbf{Feature (s) for Forecasting} & \textbf{Regression} & \textbf{reg-ARIMA} & \textbf{XGB} & \textbf{LSTMs}\\
    \hline
    \multirow{4}{*}{1} & $X_{o}$, $X_{d}$, $X_{s}$ & 26.80 & 26.33 & 22.13 & 13.14\\
    & $p-X_{u}$ & 27.13 & 26.49 & 23.20 & 17.27\\
    & $p-X_{t}$ & 26.73 & 26.05 & 23.62 & 16.90\\
    & $p-X_{de}$ & 26.50 & 22.46 & 21.81 & 16.89\\
    \hline
    \multirow{4}{*}{2} & $X_{o}$, $X_{d}$, $X_{s}$ & 34.61 & 18.00 & 17.43 & 15.41\\
    & $p-X_{u}$ & 33.57 & 21.20 & 19.31 & 17.26\\
    & $p-X_{t}$ & 32.54 & 17.96 & 18.76 & 17.37\\
    & $p-X_{de}$ & 32.47 & 18.02 & 17.98 & 17.41\\
    \hline
    \multirow{4}{*}{3} & $X_{o}$, $X_{d}$, $X_{s}$ & 49.40 & 36.54 & 33.40 & 31.35\\
    & $p-X_{u}$ & 50.74 & 42.03 & 39.00 & 32.35\\
    & $p-X_{t}$ & 49.35 & 40.06 & 46.68 & 32.45\\
    & $p-X_{de}$ & 48.73 & 38.19 & 42.91 & 32.16\\
    \hline
    \end{tabular}
    \caption{MAPE of EV consumption forecasts for all clusters}
    \label{table:mape_ev}
\end{table}

\textcolor{black}{As discussed in section \ref{lstm}, we compared several LSTM architectures and found that no specific architecture was suitable for all the clusters. However, a bidirectional vanilla architecture consistently featured among the top three models for all the clusters within the given sample size.}

\medskip

\textcolor{black}{We leverage the models of EV users and consumption to generate forecasts under different scenarios of EV penetration by varying distribution of EV types in a specimen distribution network in the UK. The forecasts so generated, therefore, capture the stochasticity that is intrinsic to real-world scenarios of EV penetration and distribution of EV types, and hence, are a faithful representation of the real-world energy consumption as different EV types gradually penetrate the automobile market. Eventually, we use the forecasts to evaluate the impact of EV charging at different levels of control under two controlling schemes, consumption control and user control. A detailed discussion is presented in section \ref{impact}.}

\section{Impact Evaluation on Distribution Transformers}
\label{impact}
\textcolor{black}{Smart or controlled charging of EVs is widely argued as a solution to offset the impact of EV penetration on both transmission and distribution networks. However, to what extent a centralized controlled charging is viable is an open question. In fact, different types of control can be considered; for example, from an EV user's point of view, a complete control on charging may not be acceptable, while some degree of freedom in EV charging even during peak hours, although costly, may be more preferable. Such a control policy will have massive revenue and generation ramifications. On the other hand, from a purely distribution-capacity point of view, a control policy that restricts the numbers of users to charge during peak hours is of greater benefit. In the following discussion, we dub these two policies as consumption and user control policies and study the impacts on a specimen distribution transformer.}

\medskip

\textcolor{black}{We consider a local distribution network in the UK as discussed in \cite{stephen2013enhanced} to evaluate the impact on a distribution transformer under stochastic scenarios of EV charging. The network incorporates a 500 kVA, 11kV/433V distribution transformer connected to four feeders (\(f_i, \, i \in \{1, 2, 3, 4\})\), with each feeder supplying to 96 households. We can reasonably assume that the total capacity of the transformer can be uniformly distributed across the four feeders, that is each feeder has a capacity of 125 kVA, indicated as \textit{f-cap} in the plots. In our EV trial data, the share of EVs between clusters 1 and 3 was 70.6\% and that between clusters 2 and 3 was 72.2\%, indicating that the EV owners had a preference of battery capacities between 4.4 kWh and 41 kWh at the time the study was carried out. Moreover, the non-EV (base) load for each household has an upper bound of 1 kVA for profile class 1 households and reaches its peak during evening hours from 5 pm to 7 pm, referred to as the peak hours, as indicated in \cite{elexon2018}. As such, under the assumption that each household can have at most one EV, we consider the following scenarios across the four feeders.}

\begin{itemize}
    \item Households belong to profile class 1.
    \item Impact is estimated during the peak hours for all the seasons in a year.        
    \item Controlled charging can constrain either the EV consumption, uniformly across all the plugged-in EVs (consumption control), or directly the number of EVs (user control) as discussed in section \ref{contribution}.
    \item Feeder 1 has EVs only from cluster 1, feeder 2 has EVs only from cluster 2, feeder 3 has 70\% EVs from cluster 1 and the remaining EVs from cluster 3, and feeder 4 has 70\% EVs from cluster 2 and the remaining EVs from cluster 3. 
    \item EV penetration level ranges from 20\% to 100\% across all the feeders.
    \item The level of control in controlled charging varies from 0 to 80\% in steps of 20\%.  
    \item As the EVs in cluster 1 have the smallest battery capacities, we assume that the power rating of all the EVs in cluster 1 is 3.5 kW, while the power rating of all the EVs in clusters 2 and 3 is 7 kW. In addition, the power factor during EV charging is considered to be 0.98 as indicated in \cite{quiros2015statistical}.
\end{itemize}

\begin{table}[htbp]
    \footnotesize
    \centering
        \begin{tabular}{|c|c|c|c|c|}
    \hline
    \textbf{Cluster} & \textbf{Plug-ins/Users per Day} & \textbf{Fraction of Users (Peak Hours)}\\
    \hline
    1 & 1.34 & 0.38\\
    \hline
    2 & 1.25 & 0.35\\
    \hline
    3 & 1.21 & 0.34\\
    \hline
    \end{tabular}
    \caption{Fraction of users charging during peak hours per day}
    \label{table:user_frac}
\end{table}

\textcolor{black}{As discussed in section \ref{contribution}, the additional load on the distribution transformer is governed by the users per day instead of the EV owners in the locality; this can be verified from figure \ref{fig:owners} which shows that the number of people charging their EVs per day is almost always less than the actual number of people who own EVs across all the three clusters. In addition, \cite{electricnation2019finalreport} identified that approximately 28\% of EV charging plug-ins per day occurred during the peak hours. Based on the ratio of plug-ins and users per day obtained from our data, we computed the fraction of users, enumerated in table \ref{table:user_frac}, who charged their EVs during the peak hours. Hence, we compute the load during peak hours based on the fraction of users plugged-in instead of owners. We generate the EV users forecasts using the XGB models developed for the three clusters to minimize the computational budget in the impact evaluation process.}

\begin{figure}[htbp]
\begin{center}
\subfigure[Feeders 1 and 2 \label{fig:con_kva_f1f2}]{\includegraphics[width=0.495\linewidth]{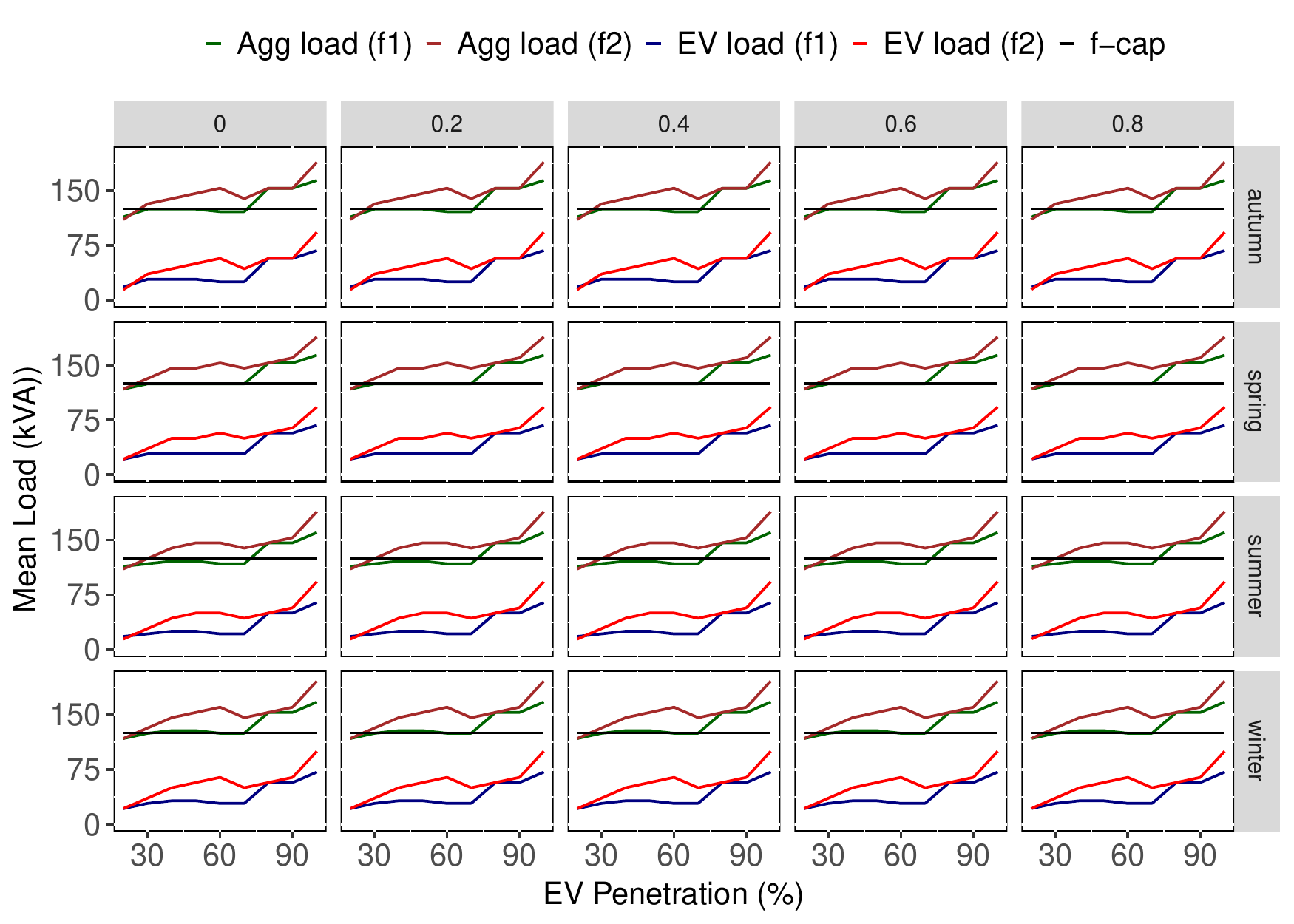}}
\subfigure[Feeders 3 and 4 \label{fig:con_kva_f3f4}]{\includegraphics[width=0.495\linewidth]{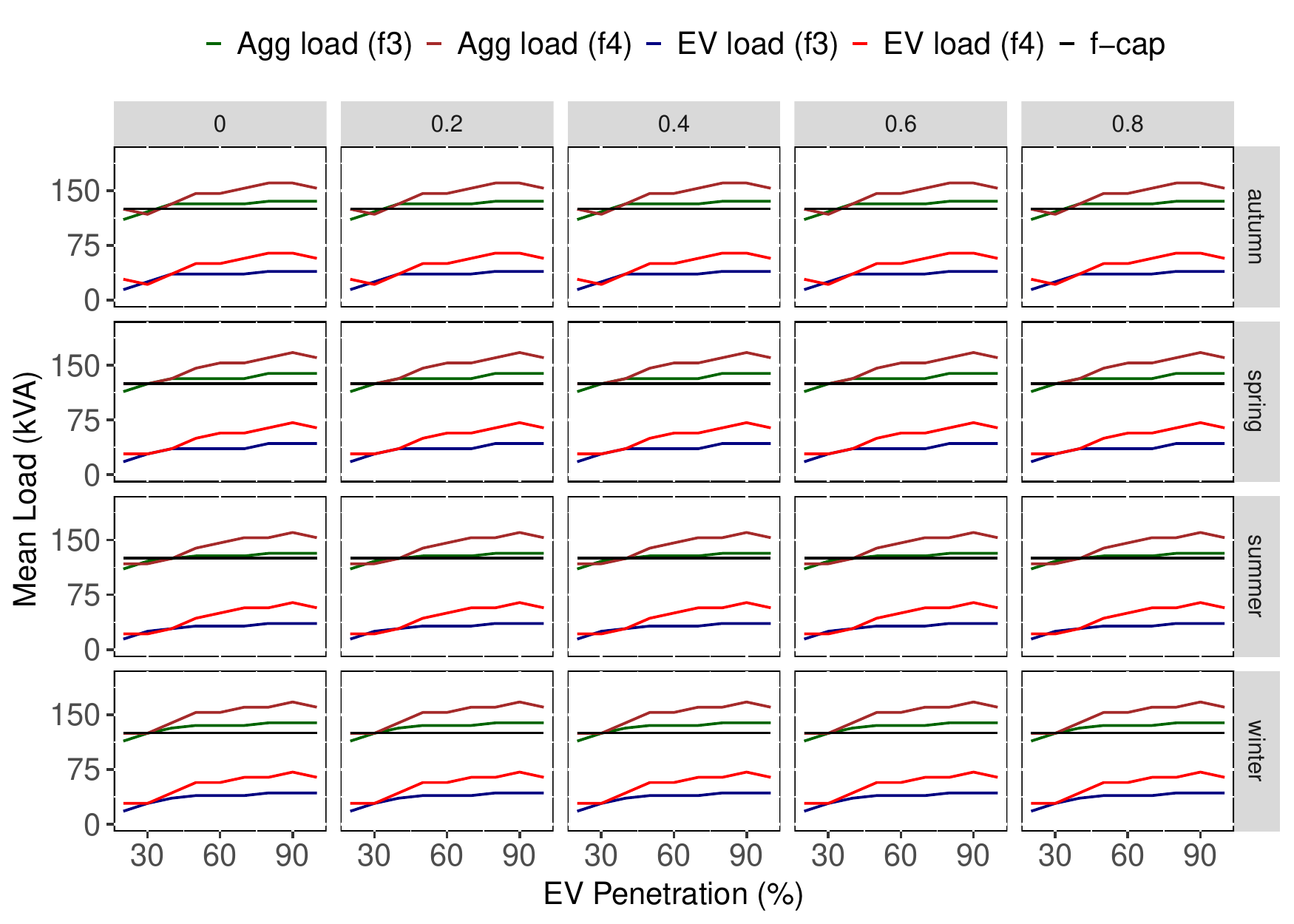}}
\caption{Mean load (kVA) due to EV charging during peak hours (consumption control)}
\label{fig:con_control_kva}
\end{center}
\end{figure}

\begin{figure}[htbp]
\begin{center}
\subfigure[Feeders 1 and 2 \label{fig:con_time_f1f2}]{\includegraphics[width=0.495\linewidth]{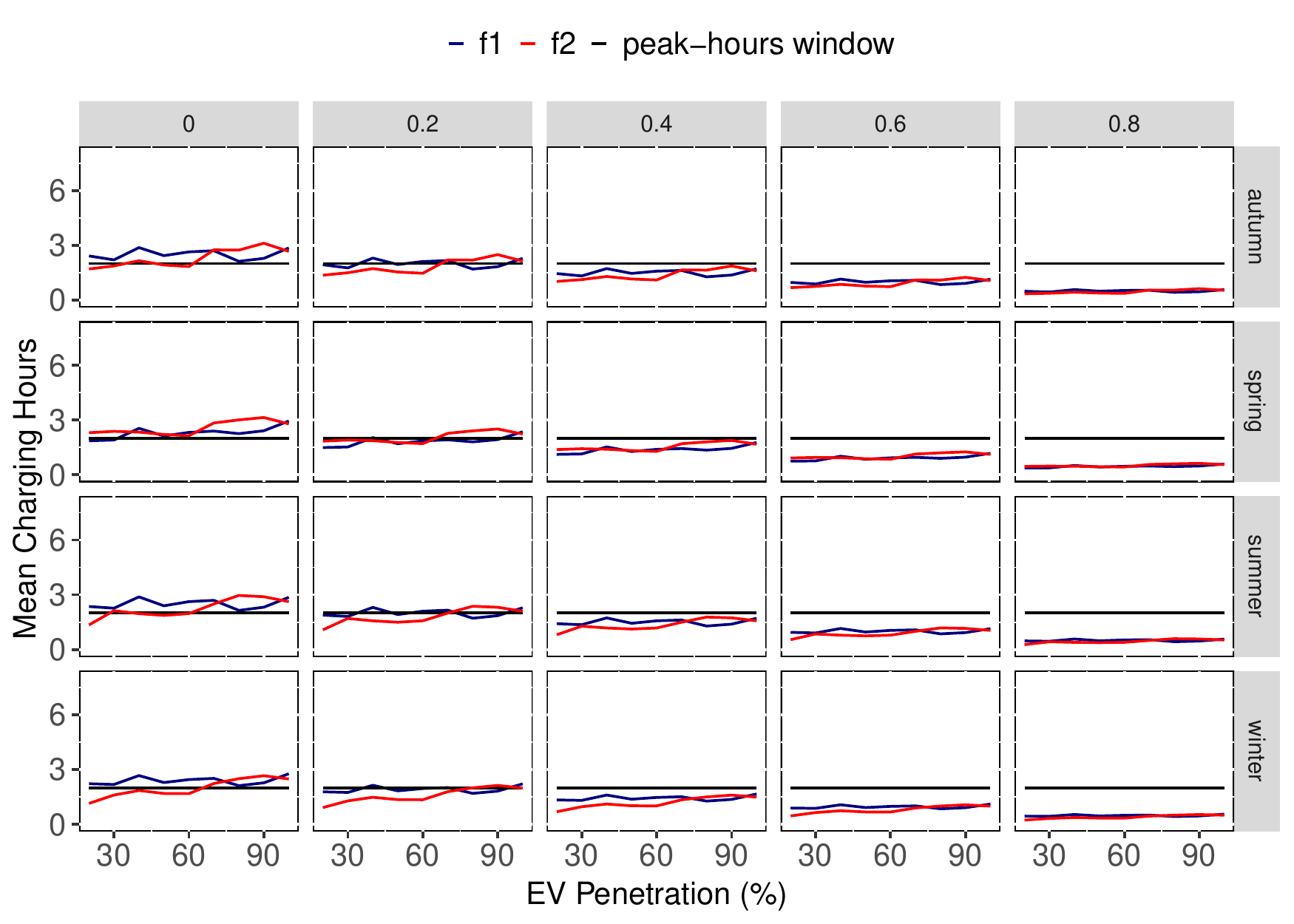}}
\subfigure[Feeders 3 and 4 \label{fig:con_time_f3f4}]{\includegraphics[width=0.495\linewidth]{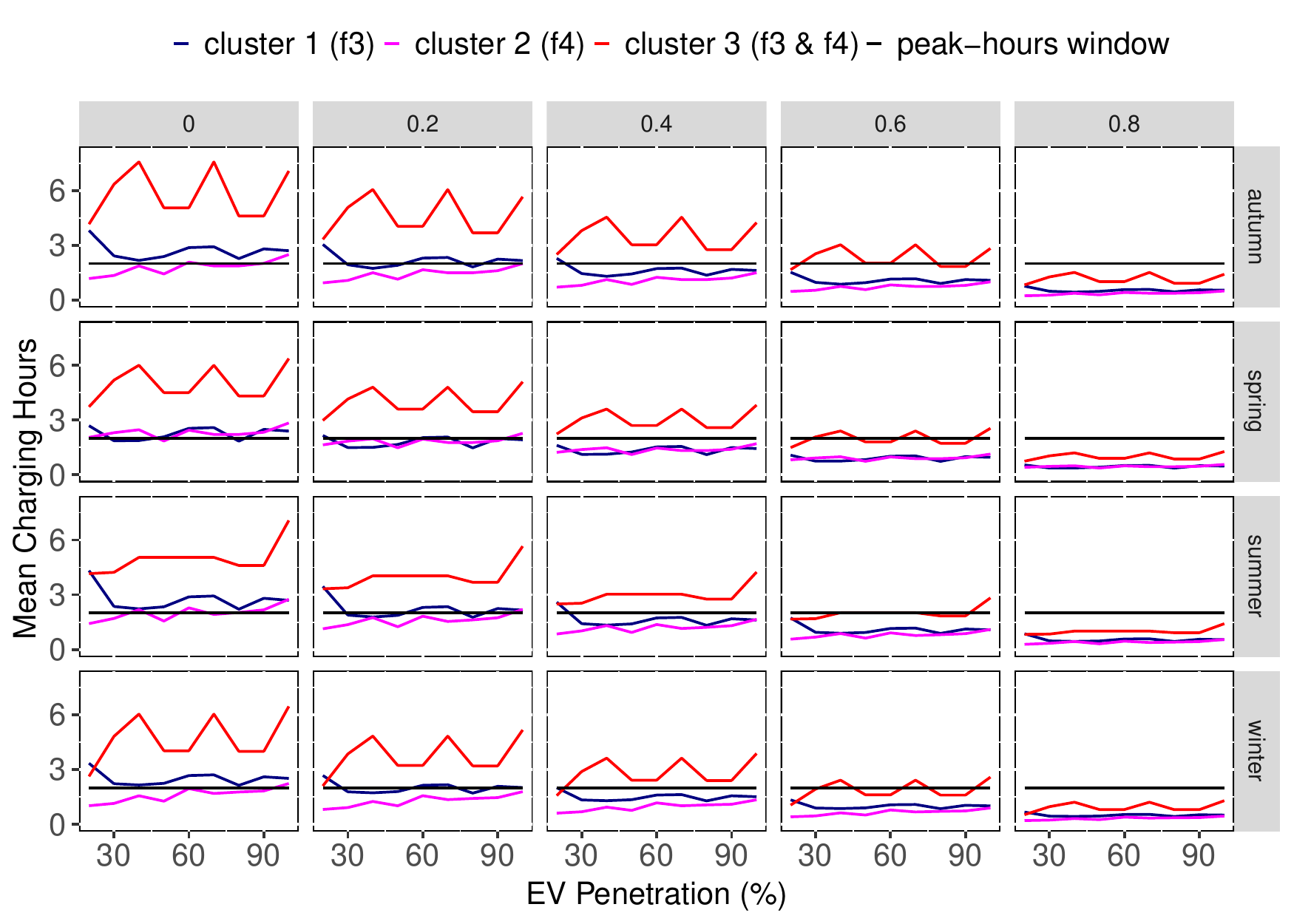}}
\caption{Mean charging duration during peak hours (consumption control)}
\label{fig:con_control_time}
\end{center}
\end{figure}

\subsection{Consumption Control vs User Control}
\label{control}
\textcolor{black}{In consumption control, all the EVs plugged-in during peak hours are allowed to consume only a fraction of their energy requirements, where the fraction is determined by the level of control, and hence, irrespective of the level of control the total connected load (sum of the EV and base loads) during peak hours, identified as \textit{Agg load} in the plots, remains the same for a given season. Figure \ref{fig:con_control_kva} shows how daily mean EV and total connected loads vary in all the feeders across all the seasons at different levels of controlled charging. We observe that the total connected load is less in feeders where there is a mix of EVs from different clusters (figures \ref{fig:con_kva_f1f2} and \ref{fig:con_kva_f3f4}). In fact, for feeders 3 and 4, the total connected load almost equals the feeder capacity (transformer capacity for each feeder) at 30\% penetration level or below for all levels of control. However, as the penetration level increases, the total connected load surpasses the feeder capacity. Figure \ref{fig:con_control_time} depicts the mean EV charging duration during peak hours. For each season, we observe that during completely uncontrolled charging, the EV charging duration exceeds the peak-hours window in feeder 1 at all penetration levels, while it remains within the peak-hours window in feeder 2 till penetration level reaches 60\% (figure \ref{fig:con_time_f1f2}). In feeders 3 and 4 (figure \ref{fig:con_time_f3f4}), EVs from cluster 3 have much higher charging duration than those from clusters 1 and 2; this follows from the fact that cluster 3 EVs have much higher battery capacities. However, as the level of control increases, the mean EV consumption during peak hours for each user decreases, thereby resulting in a decrease in charging duration.}

\begin{figure}[htbp]
\begin{center}
\subfigure[Feeders 1 and 2 \label{fig:use_kva_f1f2}]{\includegraphics[width=0.495\linewidth]{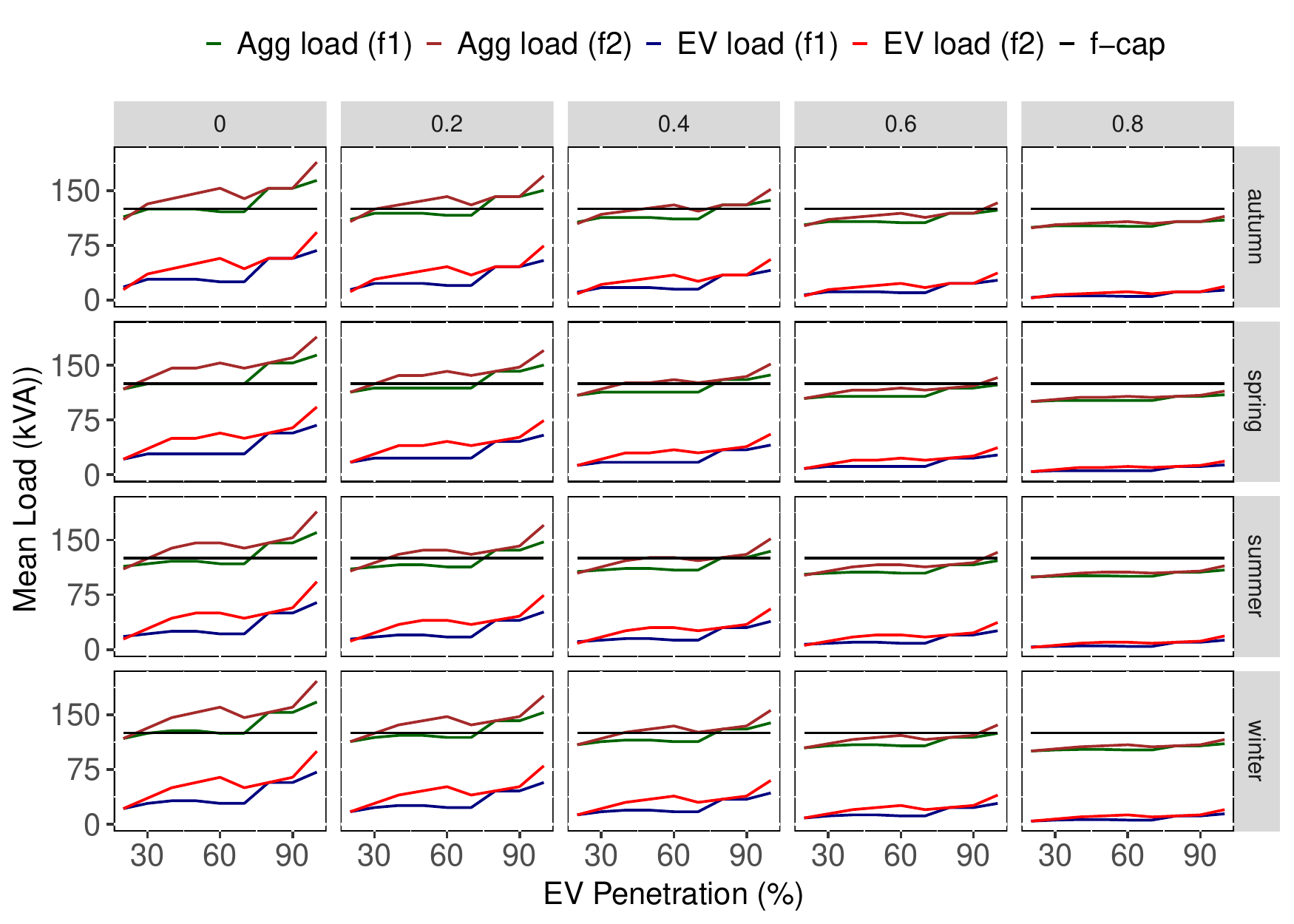}}
\subfigure[Feeders 3 and 4 \label{fig:use_kva_f3f4}]{\includegraphics[width=0.495\linewidth]{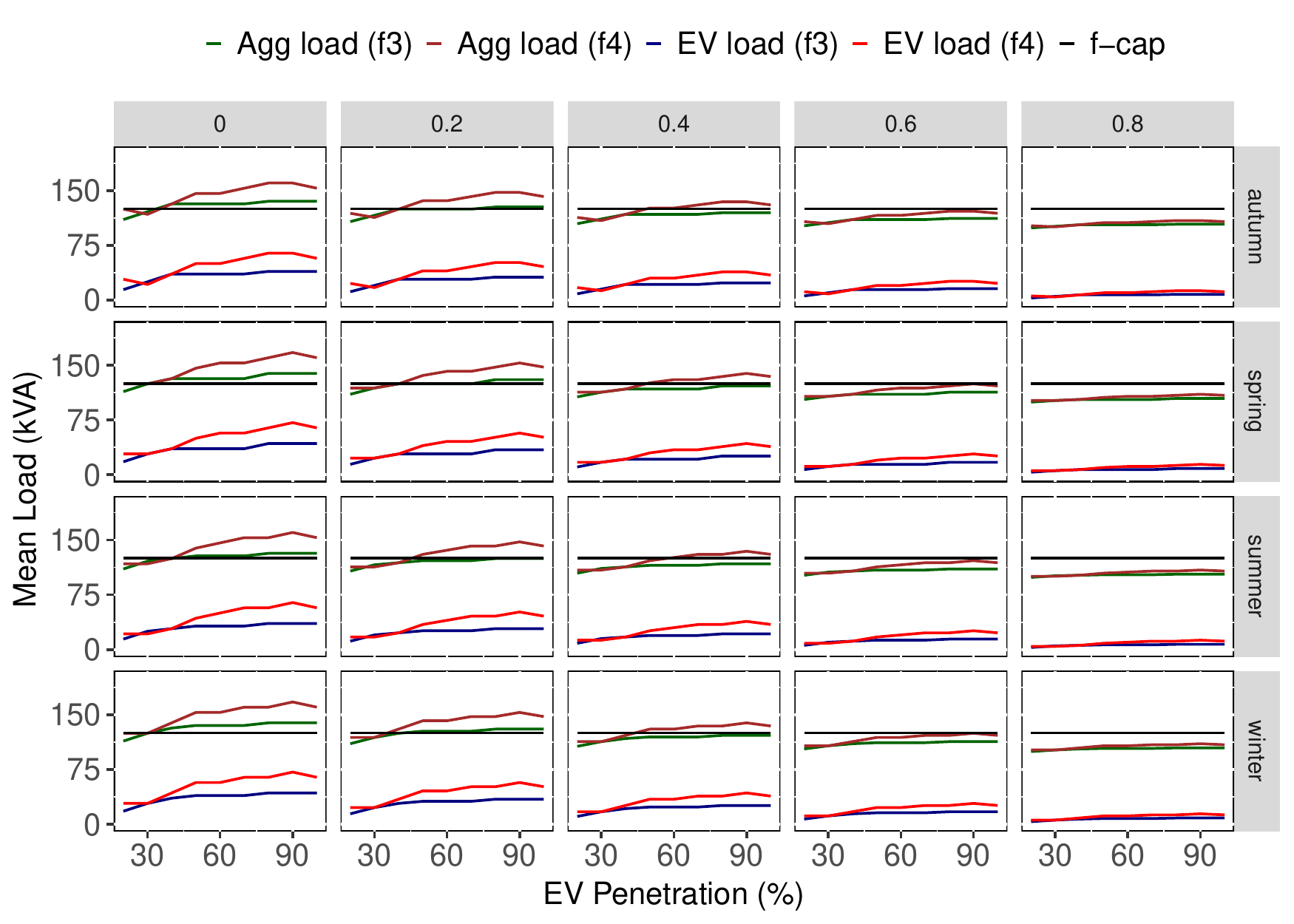}}
\caption{Mean load (kVA) due to EV charging during peak hours (user control)}
\label{fig:use_control_kva}
\end{center}
\end{figure}

\begin{figure}[htbp]
\begin{center}
\subfigure[Feeders 1 and 2 \label{fig:use_time_f1f2}]{\includegraphics[width=0.495\linewidth]{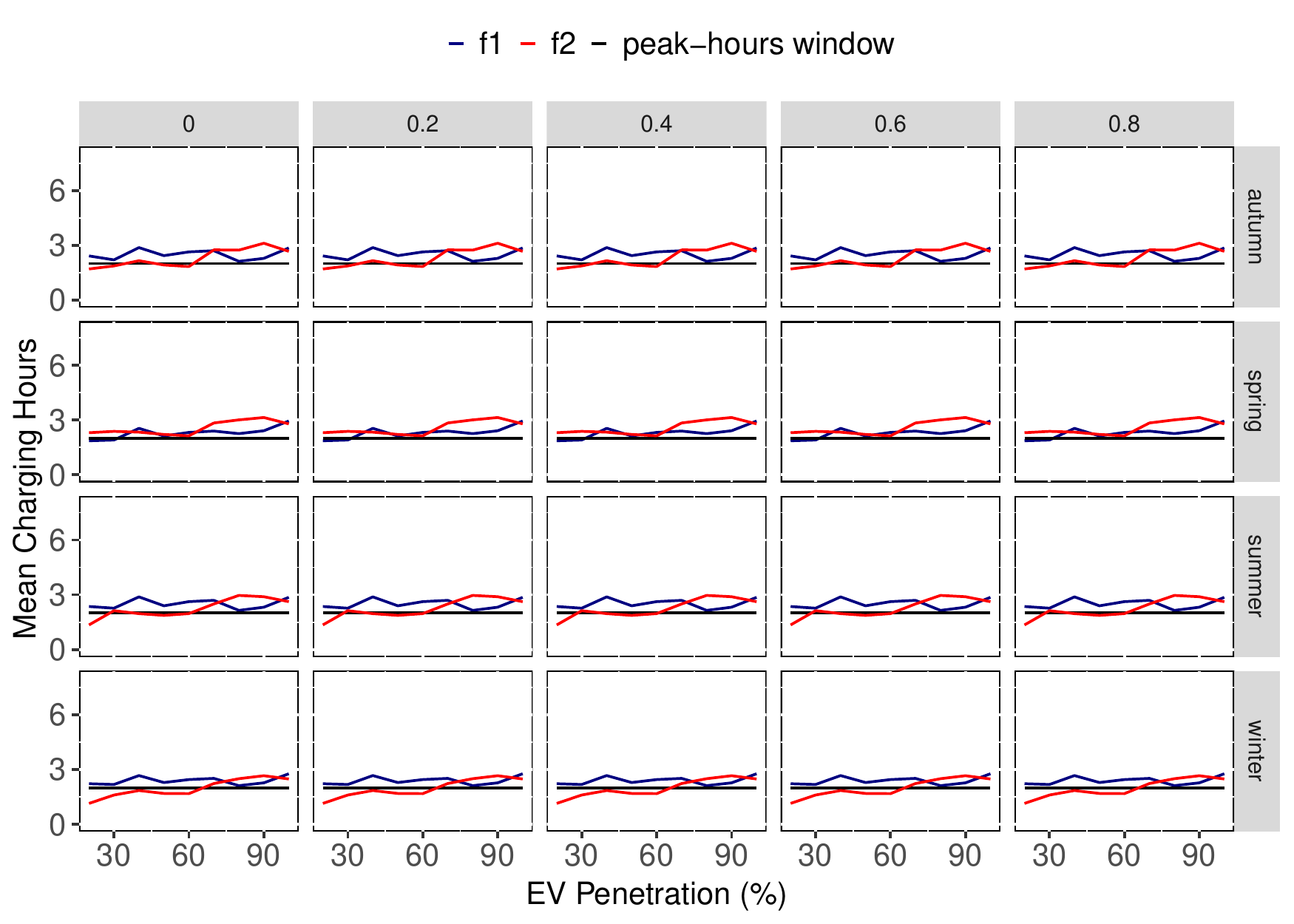}}
\subfigure[Feeders 3 and 4 \label{fig:use_time_f3f4}]{\includegraphics[width=0.495\linewidth]{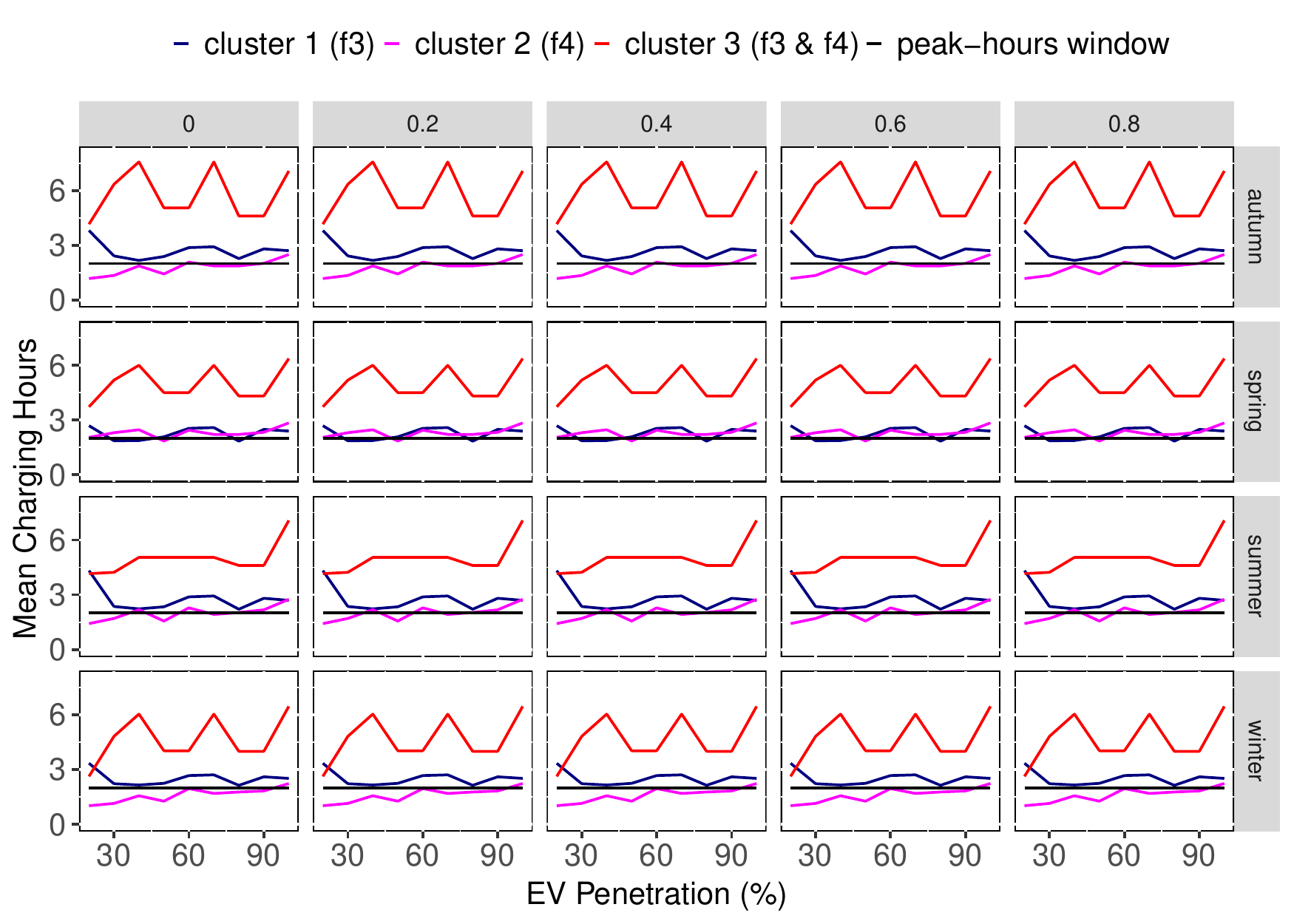}}
\caption{Mean charging duration during peak hours (user control)}
\label{fig:use_control_time}
\end{center}
\end{figure}

\textcolor{black}{In user control, only a fraction of the EVs plugged-in during peak hours, the fraction being determined by the level of control, are allowed to charge, while the charging of remaining EVs is shifted to off-peak hours and hence, as the level of control increases the total connected load during peak hours decreases for a given season. Figure \ref{fig:use_control_kva} shows how daily mean EV and total connected loads vary in all the feeders across all the seasons at different levels of controlled charging. Here also we observe that the total connected load is less in feeders where there is a mix of EVs from different clusters (figures \ref{fig:use_kva_f1f2} and \ref{fig:use_kva_f3f4}). In fact, for feeders 3 and 4, the total connected load almost equals the feeder capacity at 30\% penetration level or below at completely uncontrolled charging during peak hours. However, as the penetration level increases, the total connected load surpasses the feeder capacity. As the level of control increases, the total connected load gradually decreases and starts falling below the feeder capacity as the level of control is increased beyond 60\%. User control does not have an impact on the charging duration as the control only affects the number of users as indicated in figure \ref{fig:use_control_time}. Besides, figures \ref{fig:use_time_f1f2} and \ref{fig:use_time_f3f4} indicate that the charging duration is higher for feeders where EVs have higher battery capacities.}

\medskip

\textcolor{black}{We observed in figures \ref{fig:con_control_time} and \ref{fig:use_control_time} that the charging duration experiences a cycle of crests and troughs with increasing EV penetration, especially significant for EVs from cluster 3; this can be attributed to two reasons. Firstly, the fact that the rate at which EV users increases is higher than the rate at which EV consumption increases for small changes in EV penetration levels, thereby resulting in a net decrease in mean EV consumption per user and hence, a decrease in mean charging duration. Secondly, this effect is more pronounced in cluster 3, having largest battery capacity EVs, where forecasting accuracy is relatively lower and hence, it is conceivable that this counter-intuitive behavior of cyclic fluctuations in charging duration can also be attributed to forecasting error. Note that forecasting accuracy will more adversely impact the duration estimates than the load estimates. }

\section{Discussion and Conclusions}
\label{conclusions}
\textcolor{black}{In this study, we evaluated several models, ranging from linear statistical models such as time-series regression, to non-linear artificial neural networks (LSTMs), to forecast users who would charge their EVs per day and the resulting energy consumption caused by their EV charging. We observed that LSTMs delivered the best forecasting performance among all the algorithms considered, although XGB delivered similar performance for clusters 2 and 3 and hence, could be a possible alternative to LSTMs.} The models were developed keeping in view of the minimal information that would certainly be available to the DNOs in the future. 
However, minimal information appears to adversely affect the forecasting performance as we move towards clusters with higher battery capacities. 
In fact, statistical analysis of our data reveals that as we move towards clusters with higher battery capacities, the fraction of EV owners charging their EVs per day drops, leading to a reduced number of total charging transactions per day. This can be attributed to several factors such as the range of the vehicle (distance the vehicle can travel before needing recharging). Given the same vehicle usage (example driving style, weather, etc.), a higher capacity battery has a greater range than a smaller capacity battery does, assuming both batteries are initially charged to 100\% of their capacities. Vehicles with a greater capacity battery (higher range) are more likely to be able to complete their next journey without charging again when compared to those with a lower battery capacity, leading to lower charging frequencies. Under such a scenario, if the number of EV owners are the same in two clusters with different battery capacities, lesser number of EVs would get charged in that cluster which has higher capacity batteries as the range of vehicle would also be influential in determining the charging frequency. On a similar note, we can identify more relevant features that might be influential in explaining the variability in EV users and consumption.

\medskip

\textcolor{black}{While evaluating the effect of EV charging on a distribution transformer, we observed that user control is a more effective approach than consumption control to limit the additional bulk load caused by EV charging during peak hours as increasing the level of control decreases the bulk load only in case of user control. In fact, if the level of control is increased beyond 60\%, then the total connected load remains within the feeder capacity. On the contrary, consumption control is found to be more effective if the DNOs wish to limit the charging duration of EVs within the peak-hours window. For low to mid-ranged battery capacity EVs, any level of control beyond 40\% contains the charging duration within the peak-hours window, while for higher-ranged battery capacity EVs, charging duration gradually falls within the desired limits beyond 60\% level of control. We also observed that the total connected load on feeders with a mixed distribution of EVs from different clusters remains significantly less than the load on feeders with EVs from purely low or mid-ranged clusters, indicating that networks with EVs having low to mid-ranged battery capacities would generate higher additional load due to EV charging and hence, DNOs should lay more emphasis on such networks than on similarly structured networks with EVs having relatively high-ranged battery capacities if limiting the bulk load is of prime concern. However, if limiting the EV charging duration is the major concern, then focus should be on feeders with EVs having high-ranged battery capacities than on feeders with EVs having low to mid-ranged battery capacities. In a real-world scenario, DNOs might need to restrict not only the additional bulk load due to EV charging but also the charging duration itself, suggesting that a mixed strategy involving both consumption and user controls at a suitable level of control would be desirable under such circumstances.}

\medskip


\textcolor{black}{In this study, the framework we consider is that of minimal information that can be used to forecast EV users and consumption. Within the data available for our study, this information is EV ownership, day of the week and season of the year. However, there is a scope to collect additional information which is generally available with city councils and DNOs such as socio-demographic information of EV owners. It is conceivable that different socio-economic profiles would have different charging patterns. It would be interesting to explore the models which take these into account in future research endeavors. To this end, a future study can combine relevant choice models which explain EV charging preferences with forecast models. In addition, significant research also focused on approaches that concern a centralized controlled charging environment for EVs, assuming access to perfect information about specific attributes such as vehicle availability and state of charge. Most importantly, almost all studies implicitly assumed all EV users are happy for their charging to be controlled. However, there is little research in evaluating the feasibility of this strong assumption. In fact, as we pointed out earlier, reports from past EV trials indicated that a sizeable proportion of the EV users may not prefer completely controlled charging. Even if a certain degree of controlled charging is acceptable to the EV users, it boils down to the question of who would control the charging in the first place. It is possible that while a specific control policy may be preferred by the utility company and EV users, it may not be acceptable to the DNO and vice-versa. As such, we pointed out two control policies that naturally arise: consumption and user controls. Using our empirical study, we showed that both policies come with a trade-off between the load on transformer and the charging duration. We found that from DNO's point of view, user control would be more attractive. It is possible that from EV users' point of view, consumption control may be more acceptable. Further research is necessitated to achieve a sustainable and efficient controlled charging environment that takes aspects from both control policies and blends them to realize optimal operation of the distribution network.}

\section{Acknowledgement}
Rahul Roy's work, as part of his M.Sc. (by Research) at Lancaster University, was sponsored by grants from the
European Regional Development Fund, EA Technology, and Lancaster University, facilitated by the Centre for Global Eco-Innovation, Lancaster University. Trivikram Dokka acknowledges the support of EPRSRC Decarbon8 funded by the UK Research and Innovation, EP/S032002/1. \textcolor{black}{The authors express their gratitude to Dr. Christopher Kirkbride (Lancaster University) and Dr. Florian Dost (The University of Manchester) for their constructive appraisal of the study.} 

\bibliographystyle{apalike}
\bibliography{main}


\appendix

\section{MAPE of p-feature Forecasts}
\label{p-features}
\textcolor{black}{Tables \ref{table:p-users}, \ref{table:p-trans}, and \ref{table:p-demand} enumerate the MAPE values of the p-feature forecasts.} \textcolor{black}{Since $p-X_{t}$ can be forecast using $X_{o}$, $X_{d}$, and $X_{s}$ or $p-X_{u}$, there are 2 possible set of input features to forecast $p-X_{t}$. Similarly, $p-X_{de}$ can be forecast using $X_{o}$, $X_{d}$, and $X_{s}$ or $p-X_{u}$ and as such, there are 2 possible set of input features to forecast $p-X_{de}$. It is important to note that since $X_{de}$ was obtained by linear transforming $X_{t}$, we would not use $p-X_{t}$ to forecast $X_{de}$.}

\begin{table}[htbp]
    \footnotesize
    \centering
        \begin{tabular}{|c|c|c|c|c|c|}
    \hline
    \textbf{Feature (s)} & \textbf{Cluster} & \textbf{Regression} & \textbf{reg-ARIMA} & \textbf{XGB} & \textbf{LSTMs}\\
    \hline
    \multirow{3}{*}{$X_{o}$, $X_{d}$, $X_{s}$} & 1 & 19.38 & 22.00 & 18.55 & 12.01\\
    & 2 & 21.99 & 14.42 & 13.60 & 11.44\\
    & 3 & 29.42 & 28.55 & 23.93 & 23.87\\
    \hline
    \end{tabular}
    \caption{MAPE of p-users ($p-X_{u}$) forecasts}
    \label{table:p-users}
\end{table}

\begin{table}[htbp]
    \footnotesize
    \centering
        \begin{tabular}{|c|c|c|c|c|c|}
    \hline
    \textbf{Feature (s)} & \textbf{Cluster} & \textbf{Regression} & \textbf{reg-ARIMA} & \textbf{XGB} & \textbf{LSTMs}\\
    \hline
    \multirow{3}{*}{$X_{o}$, $X_{d}$, $X_{s}$} & 1 & 23.35 & 23.54 & 20.16 & 13.33\\
    & 2 & 25.20 & 13.85 & 13.91 & 12.16\\
    & 3 & 32.59 & 31.04 & 28.24 & 28.17\\
    \hline
    \multirow{3}{*}{$p-X_{u}$} & 1 & 23.05 & 23.10 & 20.71 & 16.24\\
    & 2 & 25.67 & 15.04 & 15.31 & 13.46\\
    & 3 & 32.49 & 30.92 & 28.52 & 27.89\\
    \hline
    \end{tabular}
    \caption{MAPE of p-trans ($p-X_{t}$) forecasts}
    \label{table:p-trans}
\end{table}

\begin{table}[htbp]
    \footnotesize
    \centering
        \begin{tabular}{|c|c|c|c|c|c|}
    \hline
    \textbf{Feature (s)} & \textbf{Cluster} & \textbf{Regression} & \textbf{reg-ARIMA} & \textbf{XGB} & \textbf{LSTMs}\\
    \hline
    \multirow{3}{*}{$X_{o}$, $X_{d}$, $X_{s}$} & 1 & 23.52 & 24.36 & 20.47 & 13.96\\
    & 2 & 25.45 & 14.31 & 13.86 & 12.58\\
    & 3 & 32.33 & 31.39 & 28.08 & 27.55\\
    \hline
    \multirow{3}{*}{$p-X_{u}$} & 1 & 23.44 & 23.99 & 20.71 & 17.16\\
    & 2 & 25.92 & 16.51 & 15.13 & 13.71\\
    & 3 & 32.05 & 29.70 & 28.55 & 27.89\\
    \hline
    \end{tabular}
    \caption{MAPE of p-demand ($p-X_{de}$) forecasts}
    \label{table:p-demand}
\end{table}

\section{Flow Diagram: The Nested Modeling Approach}
\label{nested_plot}
\begin{figure}[htbp]
\begin{center}
\subfigure[Step 1 \label{fig:nm_1}]{\includegraphics[width=0.8\linewidth]{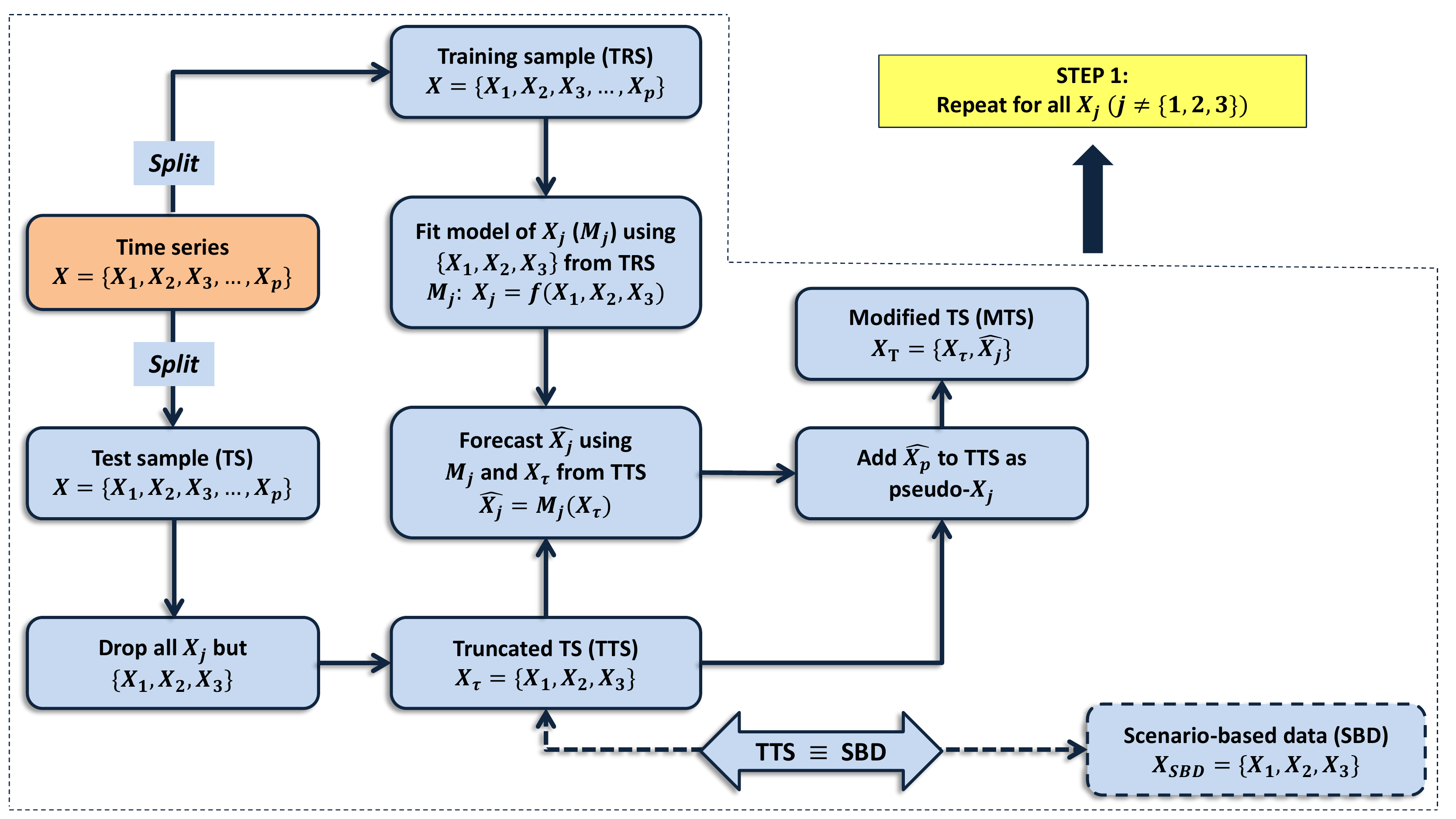}}
\subfigure[Step 2 \label{fig:nm_2}]{\includegraphics[width=0.8\linewidth]{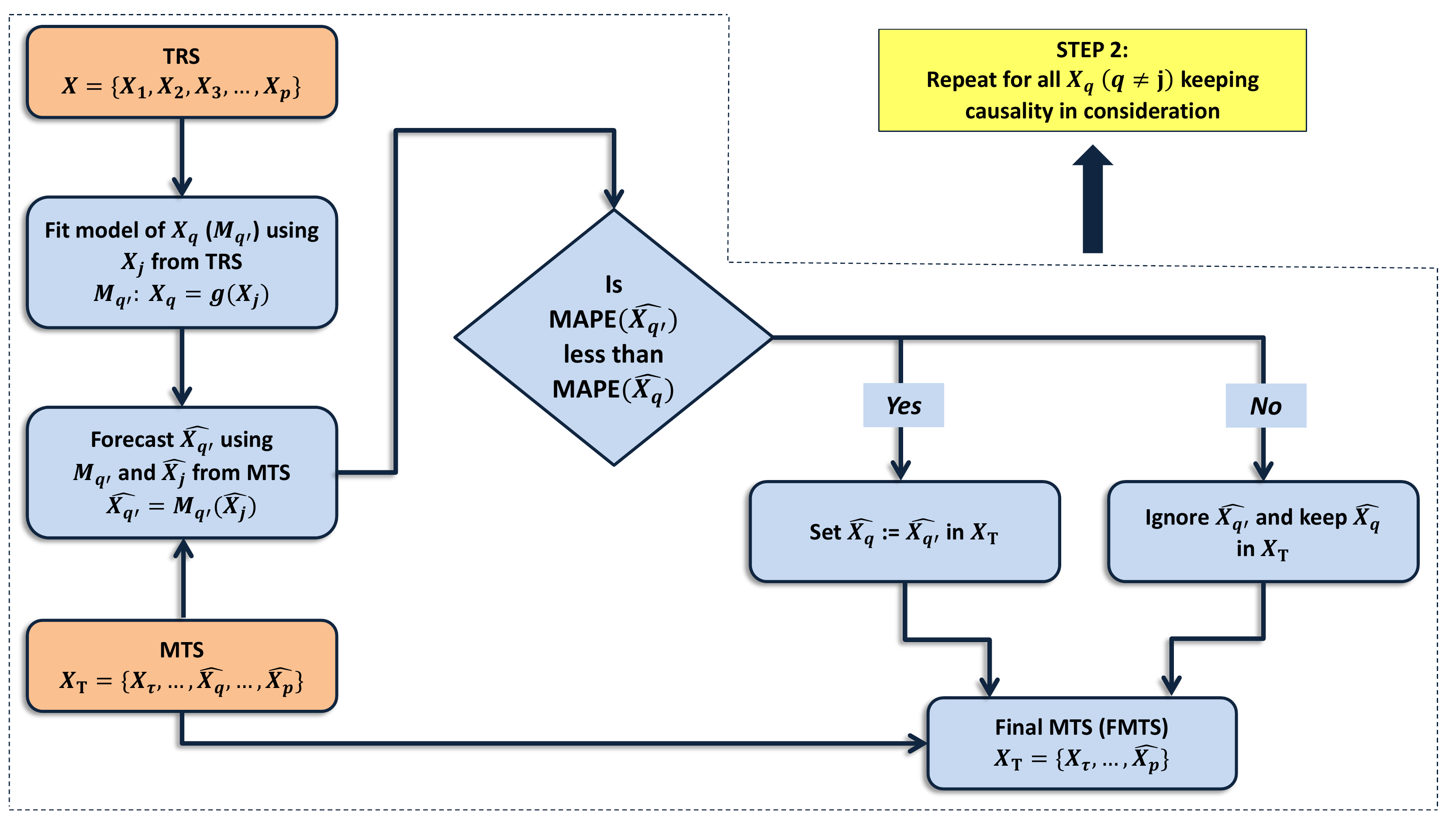}}
\subfigure[Step 3 \label{fig:nm_3}]{\includegraphics[width=0.8\linewidth]{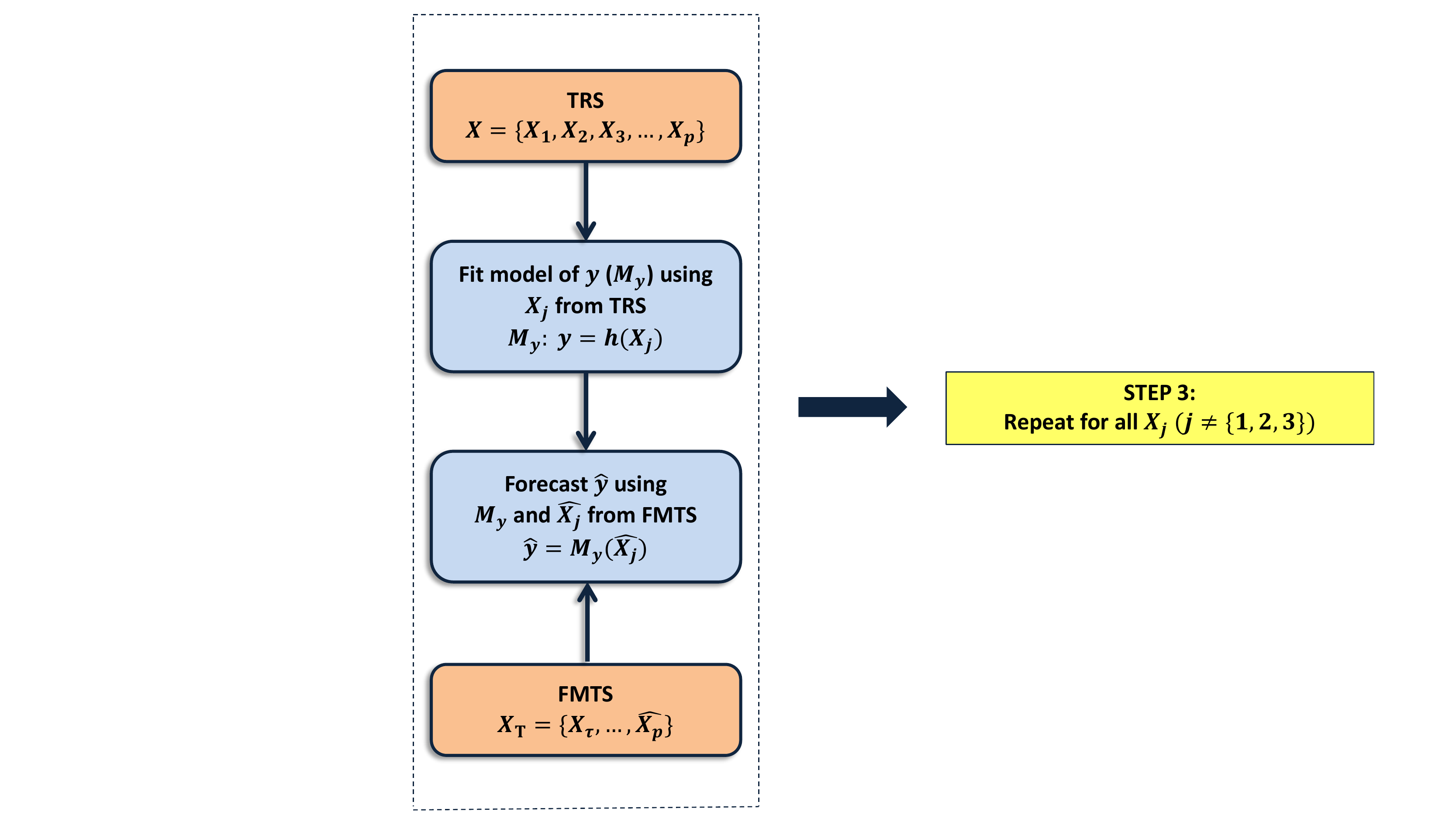}}
\caption{The nested modeling approach}
\label{fig:nested}
\end{center}
\end{figure}

Figure \ref{fig:nested} delineates the nested modeling approach, discussed in section \ref{nested modeling}, via a flow diagram. 

\end{document}